\documentclass{jfm}
\usepackage{graphicx}
\usepackage{epstopdf, epsfig}
\usepackage{natbib}
\usepackage{amsmath} 
\usepackage{amssymb}

\usepackage{wrapfig}
\usepackage{caption}
\usepackage{subcaption}
\usepackage{color}
\usepackage{bm}
\usepackage{placeins}
\usepackage{multirow,booktabs}
\newcommand{\la}{\langle}
\newcommand{\ra}{\rangle}

\newcommand\uo{{\bf u}_{_{2D}}}
\newcommand\tuo{\tilde{\bf u}_{_{2D}}}
\newcommand\uq{{\bf  u}_q }

\newcommand\vq{{\bf  v}_q }
\newcommand\tvq{\tilde{\bf v}_q }

\newcommand\QTC{Q_{_{3D}} }
\newcommand\QDC{Q_{_{2D}} }

\newcommand\eop{ \epsilon_{_{2D}}^+ }
\newcommand\eom{ \epsilon_{_{2D}}^- }
\newcommand\eqp{ \epsilon_{_{3D}}^+ }

\newcommand{\beq}{\begin{equation}}
\newcommand{\eeq}{\end{equation}}

\shorttitle{Critical Transitions in Thin Layer Turbulence}
\shortauthor{S.J. Benavides \& A. Alexakis}

\title{Critical Transitions in Thin Layer Turbulence }

\author{Santiago Jose Benavides\aff{1} \& Alexandros Alexakis\aff{2}
	\corresp{\email{alexakis@lps.ens.fr}} }

\affiliation{
	\aff{1} Department of Earth, Atmospheric and Planetary Sciences, Massachusetts Institute of Technology, 77 Massachusetts Ave., Cambridge, MA 02139-4307, USA \aff{2}  Laboratoire de Physique Statistique, \'{E}cole Normale Sup\'{e}rieure, CNRS, Universit\'{e} Pierre et Marie Curie, Universit\'{e} Paris Diderot, Paris 75005, France }

\begin{document}

\maketitle

\begin{abstract}

We investigate a model of thin layer turbulence that follows the evolution of 
the two-dimensional motions $\uo (x,y)$ along the horizontal directions $(x,y)$ 
coupled to a single Fourier mode along the vertical direction ($z$) of the form $\uq(x, y, z)=[v_x(x,y) \sin(qz), v_y(x,y)\sin(qz), v_z(x,y)\cos(qz)\, ]$,
reducing thus the system to two coupled, two-dimensional equations. 
The reduced dimensionality of the model allows a thorough investigation of the transition 
from a forward to an inverse cascade of energy as the thickness of the layer $H=\pi/q$ is varied. 
Starting from a thick layer and reducing its thickness 
it is shown that two critical heights are met 
(i)  one for which the forward unidirectional cascade (similar to three-dimensional turbulence) transitions to 
     a bidirectional cascade transferring energy to both small and large scales and 
(ii) one for which the bidirectional cascade transitions to a unidirectional 
     inverse cascade when the layer becomes very thin (similar to two-dimensional turbulence).
The two critical heights are shown to have different properties close to criticality that we are able to analyze with numerical simulations 
for a wide range of Reynolds numbers and aspect ratios.

\end{abstract}

\section{ Introduction }\label{sec:intro}         

Turbulence prevails in the universe, and its multi-scale properties
affect the global dynamics of geophysical, astrophysical, and industrial flows.  
Typically, in a turbulent flow, energy is supplied at some scale and is redistributed among scales due to the stretching of vortices 
by interactions with similar size eddies. 
%
This mechanism of energy transfer from large to small scales or vice versa is known as a forward or inverse cascade, respectively.
A prominent example of a forward cascade is met in three-dimensional (3D) hydrodynamic turbulence \citep{Frisch}.
The turbulent cascade in this case transports the energy from the large (possibly coherent) structures to small
`incoherent' scales.  An example of an inverse cascade is given by two-dimensional (2D) hydrodynamic turbulence that cascades energy to the large scales \citep{2DT}.
There are some examples, however, that have a mixed behavior, such as rapidly rotating fluids, 
conducting fluids in the presence of strong magnetic fields, flows in constrained geometry and others.
In these examples the injected energy cascades both forward and inversely in fractions that depend on the value of 
a control parameter (rotation rate/magnetic field/aspect ratio/etc.). These system thus exhibit a {\bf bidirectional
cascade}: coexistence of a forward and an inverse cascade of energy whose relative amplitudes depend on parameters of the system.

Bidirectional cascades have been observed in different physical situations both
in numerical simulations \citep{smith1996,Smith1999,2DH,B0,Rot,geophys,us1,Rot2,strat,geophys2,geophys3,geophys4,us2} and
in laboratory experiments \citep{exper2,exper1,exper3,Yarom2013,Moisi}. 
They are met in atmospheric physics, where in large scales the atmosphere acts like a two-dimensional flow 
cascading energy inversely while at the same time in small scales it acts as a three-dimensional flow 
cascading energy to the even smaller scales. The amplitude of these cascades (both inverse and forward)
have been quantified from in situ aircraft measurements in the hurricane boundary layer \citep{huricanes}.
Similar behavior has been claimed in astrophysical flows (like the atmosphere of Venus \citep{Venus} and accretion discs \citep{Accretion}) and 
industrial applications (like tokamak plasma flows \citep{PatDiamond}) either due to the thinness of the layer, fast rotation or the presence of strong magnetic fields.

This work focuses on one of the simplest setups that exhibits such a transition: turbulence in a thin layer.
By thin layer we refer to a 3D domain that extends to large distances $L$ in two (horizontal) directions 
and short distance $H$ in the third (vertical) direction. 
In such a system eddies with scales $\ell$ much larger than the layer thickness $H \ll \ell$ are 
constrained to two-dimensional dynamics while small eddies in the opposite  limit $\ell \ll H$ do not to feel this constraint
and behave like a three-dimensional flow. This system was first examined by \cite{2DH} where
it was shown that for large $Re$ the direction of the energy cascade depends on 
the ratio $Q=\ell_f/H$ of the horizontal length-scale of the forcing $\ell_f$ to the layer height $H$. 
Alternatively, $Q$ can be considered as the ratio of the smallest non-zero vertical wavenumber to the forcing wavenumber.
For $Q \gg 1$ $(\ell_f \gg H)$
the energy is injected in eddies that fall in the first regime (2D) and therefore the cascade is inverse,
while for $Q\ll1$ $(\ell_f \ll H)$ 
the energy is injected in eddies that fall in the second regime (3D) and therefore the cascade is forward.
At intermediate values however the system displays a bidirectional cascade \citep{2DH}.

In this work we will focus on the exact way that the system transitions from a unidirectional to a bidirectional  cascade.
There are three possible scenarios for such a transition  
(a) the transition happens in a smooth way.  
In this case the amplitude of the inverse or forward cascade decreases smoothly as $Q$ is varied
(possibly as a power-law)
and therefore the inverse cascade becomes zero only at the limit  $Q \to 0$ 
while the forward cascade becomes zero only at the opposite limit $Q \to \infty$.
(b) The transition happens at a critical value $Q=Q_c$ in a discontinuous way, much like a sub-critical instability or a first order phase transition,
so that the system changes abruptly from an inverse cascade to a forward cascade.
(c) the amplitude of the inverse/forward cascade decreases/increases continuously as $Q$ is increased and 
at a critical point $Q_c$ the inverse/forward cascade becomes exactly zero with discontinuous or diverging derivatives 
much like a super-critical instability or a second order phase transition. 
In the work of \cite{2DH} it was conjectured that the third scenario (c) is observed for the transition from a forward unidirectional cascade 
to a bidirectional cascade although their data did not allow to draw a precise conclusion.

\begin{figure*}                                                                                                               %
\centering{                                                                                                                   %
	\def\svgwidth{.8\textwidth}                                                                                                 %
	\large{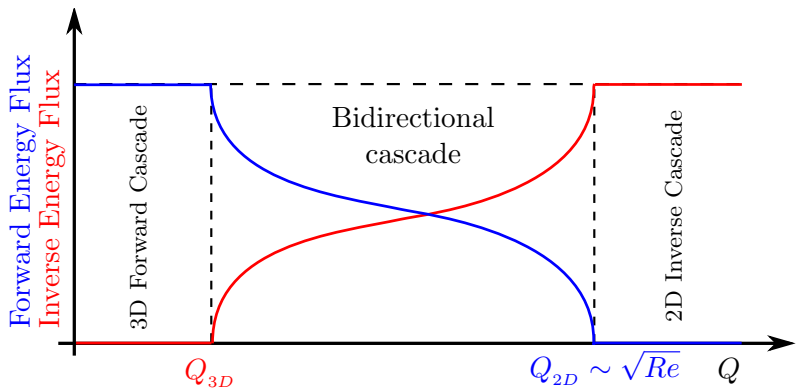}}                                                                                   %
\caption{An illustration of the expected critical transitions, corresponding to scenario (c) described in the text.           %
        $Q=\ell_f/H$ is the ratio of the horizontal length-scale of the forcing $\ell_f$ to the layer height $H$.}            %
\label{scenario}                                                                                                              %
\end{figure*}                                                                                                                 %

For the case of a critical transition there a few remarks that we need to make. 
These systems transition from one turbulent state that cascades energy inversely, say, to the
large scales to a different turbulent state that cascades energy forward to the small scales. Turbulence
is thus always present! This makes the discussed transition far different from the traditional scenarios of
transition from a laminar to a turbulent state. Turbulent fluctuations are always present and
constitute an integral part of the mechanism for the transition, much like how thermal fluctuations
play a determinant role in equilibrium phase transitions close to criticality.

For turbulence in thin layers we expect two critical values of $Q$ to exist.
The first, $\QTC$, marks the transition from purely forward to a bidirectional cascade, whereas the second one, $\QDC$, marks the transition from a bidirectional cascade to a purely inverse cascade.
In more detail, for flows of large $Re$ and
for layers sufficiently thick $Q\ll1$ we expect that all energy cascades towards the small scales 
and no energy to the large scales. As $Q$ is increased, and thus the layer made thinner, a critical value $\QTC$ will be met for which 
the appearance of an inverse cascade will begin in coexistence with the forward cascade.
This marks the beginning of the bidirectional cascade. As $Q$ is increased further the forward cascade
decreases while the inverse cascade increases. We then expect a second critical height $\QDC$ where the forward cascade
becomes zero and all the energy cascades to the large scales. Further increase of $Q$ will not alter this behavior.
The bidirectional cascades then exist in the range $\QTC<Q<\QDC$. An illustration of this expectation is shown in Figure \ref{scenario}. 

The purpose of the present work is to focus on the two critical points: unravel 
their statistical behavior as well as study the mechanisms involved close to criticality. 
Performing such a study with direct numerical simulations  of the three-dimensional Navier-Stokes equation is computationally costly 
due to the high degree of resolution required in order to have both large enough Reynolds number 
so that the flow is turbulent and large  enough scale separation $L\gg \ell_f$ so that an inverse cascade develops.
To overcome this difficulty we will instead focus on a model of the Navier-Stokes equation. 
In our model we will keep a minimal description for the vertical direction by performing a drastic 
Galerkin truncation in the vertical direction by keeping only two modes $\uo(t,x,y)$ and $\uq(t,x,y,z)$. 
The first mode corresponds to purely two-dimensional motions and depends only on the horizontal directions $(x,y)$, 
while the second mode corresponds to a flow whose vertical structure is proportional to $\sin(q z)$ or $\cos(q z)$
and has arbitrary dependence in the horizontal directions.

The rest of this paper is structured as follows. Section \ref{sec:Model} describes our model of thin layer turbulence, as well as our methodology, explaining our measures of the inverse and forward energy cascades (acting as order parameters) as well as the simulation details themselves. Section \ref{sec:Q3D} presents the results and analysis of our investigation of the transition from a purely forward energy cascade to a bidirectional energy cascade. Section \ref{sec:Q2D} focuses on the other transition between a bidirectional cascade and a pure inverse cascade. Finally, in Section \ref{sec:Conc} we summarize our findings and give concluding remarks regarding the results.

\section{Model Description and Methodology }\label{sec:Model}              

We consider an incompressible flow in a thin layer of thickness $H=\pi/q$ in the vertical direction ($z$) 
and of size $2\pi L$ in the remaining two directions $(x,y)$ with  free slip boundary conditions $\partial_z u_x =  \partial_z u_y = u_z = 0$ at $z=\pm H/2$.
The the flow velocity $\bf u$ is governed by the Navier-Stokes equations:
\begin{align} \label{NS}
\partial_t {\bf u } + {\bf u}\cdot \nabla {\bf u } &= -{\nabla P} + \nu \Delta {\bf u} + {\bf F} \\
\nabla \cdot {\bf u} &= 0,
\end{align}
where $\nu$ is the kinematic viscosity, $P$ is the pressure, and $\bf F$ is a two-dimensional forcing (varying only along $x$ and $y$)
that acts at a particular horizontal length-scale scale $\ell_f$. 

As discussed in Section \ref{sec:intro}, in this work we do not solve for the full system requiring three-dimensional numerical simulations, 
but we rather focus on a model that is obtained by a severe Galerkin-truncation in the vertical direction such that only two modes 
are kept. The first mode $\uo$ has two components and corresponds to a purely two-dimensional mode. It can be obtained by vertical averaging
$\bf u$. The incompressibility condition allows to write $\uo$ in terms of a stream function $\psi(t,x,y)$
as indicated in Eq. \ref{modes}.
The second mode $\uq$ has all three components and a prescribed vertical dependence as given below:
\beq
		\uo (t,x,y) = \left(
		\begin{array}{c} \partial_y \psi \\ - \partial_x \psi \\ 0 \end{array} 
		\right), 
		\quad 
		\uq (t,x,y,z) = \left(
		\begin{array}{c} v_x (x,y,t)\sin(q z)\\ v_y(x,y,t)\sin(q z) \\ v_z(x,y,t)\cos(q z) \end{array} 
		\right). 
		\label{modes}
\eeq
The vector field $\uq$ satisfies free slip boundary conditions at $z=\pm H/2$ and the incompressibility condition
 $\nabla \cdot \uq=0$ that, in terms of $\vq(t,x,y) = (v_x,v_y,v_z)$, is written as $\partial_x v_x + \partial_y v_y = q v_z$. 
With this notation the truncated Navier-Stokes equations can be written as
\begin{align}
\partial_t \uo + \uo \cdot \nabla \uo &= -\overline{ \uq \cdot  \nabla \uq } - \overline {\nabla P}  +  \nu \Delta \uo  -  \mu \Delta^{-2} \uo   + \bf F, \label{thinlayer_a} \\
\partial_t \uq + \uo \cdot \nabla \uq &= -           \uq \cdot  \nabla \uo   -            \nabla p_q +  \nu \Delta \uq  , 
\label{thinlayer_b}
\end{align}
where the over-bar stands for vertical averaging $\overline{f} \equiv \frac{1}{H}\int { f }dz$ and $p_q$ is the partial pressure that guaranties the incompressibility of $\uq$.
Note that if one plugs in $\uq$ from Eq. \ref{modes} into Eq. \ref{thinlayer_b}, the vertical dependence drops out and we are left with two coupled partial differential equations that depend only on the horizontal directions $(x,y)$.
Furthermore, note that in our system we have included the hypo-dissipation term $\mu \Delta^{-2} \uo$ that is responsible for saturating the inverse cascade when present.
If such a term is absent in the presence of an inverse cascade the energy of the large scale modes will grow to very large values forming a condensate \citep{2DT,chertkov2007dynamics,condensate}
whose amplitude growth is balanced by the small viscous forces. We would like to avoid this situation so in all our runs $\mu$ is tuned so that the largest scale is sufficiently suppressed.

In the absence of forcing and dissipation terms this system conserves the total energy of the flow given by
\beq
E=\frac{1}{2}\langle |\uo|^2 + |\uq|^2 \rangle = \frac{1}{2} \langle | \nabla \psi |^2 \rangle + \frac{1}{4} \langle v_x^2 +v_y^2 +v_z^2 \rangle,
\eeq
where angular brackets $\langle \cdot \rangle$ stand for spatial average. For $\uq=0$ one recovers the two-dimensional Navier-Stokes equation in which case the enstrophy 
$\Omega = \frac{1}{2} \langle |\nabla \times \uo|^2 \rangle=\frac{1}{2} \langle |\Delta \psi|^2 \rangle$ is also conserved.
In the presence of forcing and dissipation the system eventually reaches a steady state where the energy injected by the forcing
at the averaged rate $I=\langle {\bf F}\cdot \uo \rangle_{_T}$
is balanced by the dissipation rates 
$\eqp$, $\eop$, and $\eom$ due to viscous and hypo-viscous forces defined as:
\begin{equation}\label{e3d}                                             %
\eop = \nu \la |\nabla \times \uo|^2 \ra_{_T} , \quad                   %
\eqp= \nu \la | \nabla \times \uq |^2 \ra_{_T}  ,\quad                  %
\eom = \mu \la |  \Delta^{-1} \uo  |^2 \ra_{_T},                         %
\end{equation}                                                          %
where the brackets $\la \cdot \ra_{_T} $ indicate space and time average.
The dissipation rate $\eop$ measures the rate energy is dissipated at the small scales
of the 2D velocity field $\uo$, while $\eqp$  measures the rate energy is dissipated
by the 3D field $\uq$. Finally $\eom$  measures the rate energy is dissipated 
at the large scales of the 2D velocity field.
At steady state a balance is reached and we obtain:
\beq                                                                                 
I = \eop + \eqp + \eom.                                                   
\eeq                                                                                 
Thus, with a large enough scale separation $L \gg \ell_f$,
 the ratio $\eom/I$ provides us with a measure of what fraction of the energy
injected is cascading to the large scales. 
On the other hand,
at high enough Reynolds numbers,
 the ratios $ \eop/I$
and $\eqp/I$ provide us with the fraction of the energy that cascades to the small 
scales.

Two different forcing functions were used in this study. The first was a deterministic time-independent forcing explicitly given by
\beq \label{det}
 {\bf F} = F_0 [ \cos(k_f y), - \cos(k_f x), 0 ],
\eeq
where $F_0$ is the forcing amplitude and $k_f = \pi / \ell_f$ is the forcing wavenumber.
This forcing, although more physical than our second choice,
it does not inject energy at a constant rate since the injection rate depends on the particular flow realization.

For a better control of the energy injection rate in our system
a second type of forcing was used that was designed to inject energy at a given shell of wavenumbers of modulus $k_f$ at a constant rate at each instant of time
and for every realization. 
It is written as 
\beq 
{\bf F} = I_0 \,  \sum_{\bf k} \frac{\tuo ({\bf k}) e^{i \bf k\cdot x}}{\sum_{\bf k^\prime}   |\tuo({\bf k'})|^2}  + i\sum_{\bf k} \Omega_{\bf k} \tuo({\bf k}) e^{i \bf k\cdot x} \label{constI}
\eeq
where $\tuo ({\bf k})$ stands for the Fourier transform of the field $\uo$.
The sums are over all wavenumbers that satisfy $|{\bf k}|=k_f$, and
$\Omega_{\bf k}$ is a random frequency that leads to a decorrelation of the phases between the different forced modes.
The total  energy injection rate for this forcing at each instant of time and for each realization is given by $I_0$.  
This forcing thus allows us to control the energy injection rate without employing a delta correlated random forcing.

The relevant non-dimensional control parameters of our system depend on the domain geometry, dissipation parameters, and on the forcing mechanism and scale.
The ratio of the inverse layer thickness to the forcing wavenumber is given by $Q \equiv q/k_f$ and is our primary control parameter.
The relative scale separation between the forcing scale and the horizontal box size is measured by $k_f L$. We used two forms of Reynolds numbers; one defined in terms of our control parameters used to provide consistency between different runs, and another measured after the fact used in analysis of our data. 
The former Reynolds number is defined as $Re_f = F_0^{1/2}/ \nu k_f^{3/2}$ for the constant forcing amplitude runs of Eq. \ref{det} and $Re_f = I_0^{1/3}/ \nu k_f^{4/3}$ for the constant energy injection rate runs of Eq. \ref{constI}. The second form of Reynolds number uses the more classical definition $Re =u_f/k_f \nu$,
where $u_f$ is the root mean square (rms) value of the velocity at the forcing scale. 
The value of $\mu$ was always tuned so that no large scale condensate was formed and thus it was always tied to
the size of the box $k_fL$. 

Equations \ref{thinlayer_a} and \ref{thinlayer_b} were solved using a standard parallel pseudo-spectral code with a fourth order Runge-Kutta scheme for time integration
and a two-thirds dialliasing rule. More details on the parallelization can be found in \cite{gomez2005}.
All runs started from random initial conditions
and were carried out 
long enough so that a statistically steady state was reached. All measurements and averages were made at this state unless otherwise stated.
The resolutions used varied from $512^2$ to $4096^2$ grid points. 

Our goal is to investigate how the system transitions to and from a bidirectional cascade in the limit of large box size (or large scale separation)
and large Reynolds numbers. Thus, close to the points of criticality, a series of runs were performed varying the parameter $Q$
for fixed values of $k_fL$ and $Re_f$. Then the same series of runs were repeated for larger values of $k_f L$ or $Re_f$ until
we observed convergence. 
The runs were separated into three cases: A, B, and C. See Table \ref{table_sim} for a list of our cases and their corresponding parameters. Cases A and B investigated the transition from a purely forward cascade to a bidirectional cascade, with a constant forcing amplitude (Eq. \ref{det}) and a constant energy injection rate (Eq. \ref{constI}), respectively. For these runs we fixed $Re_f$ while varying $k_fL$, and measured the inverse energy cascade (which we expected to transition from being zero to being nonzero) via $\eom/I$. Finally, case C investigated the transition from a bidirectional cascade to a purely inverse energy cascade. For these runs we fixed $k_f L$, varied $Re_f$, and measured the forward energy cascade (which we expected to transition from being nonzero to zero) via $\eqp / I$.

\begin{table}
	\begin{center}
		\begin{tabular}{ c c c c c c c c c c c c }
			\hline \hline
			 Case	\quad	& 	A1  &  A2  & A3 & \quad B1 & B2 & B3 & B4 & \quad C1 & C2 & C3 & C4  \\ \hline
			 $k_f L$ \quad 	&   8   &  16  & 32 & \quad 8 & 16 & 32 & 64 & \quad 8 & 8 & 8 & 8 \\ 
			 $N$ \quad		&   512 &  1024 & 2048 & \quad 512 & 1024 & 2048 & 4096 & \quad 512 & 512 & 512 & 1024 \\ 
			 $Re_f$	\quad	&   59  &  59  &  59 & \quad 66 & 66 & 66 & 66 & \quad 44 &  59 & 88 & 177 \\
			 Forcing \quad		&   \multicolumn{3}{c}{$F_0$, Eq. \ref{det}} & \multicolumn{4}{c}{$I_0$, Eq. \ref{constI}} & 	&   \multicolumn{3}{c}{$F_0$, Eq. \ref{det}} \\ \hline \hline
		\end{tabular}
		\caption{A summary of the parameter ranges used in our simulations.}
		\label{table_sim}
	\end{center}
\end{table}\vspace{-.5cm}

\section{ Transition from a forward to a bidirectional cascade}\label{sec:Q3D} 

\begin{figure*}
	\begin{center}\vspace{0cm}
		\includegraphics[width=0.48\textwidth]{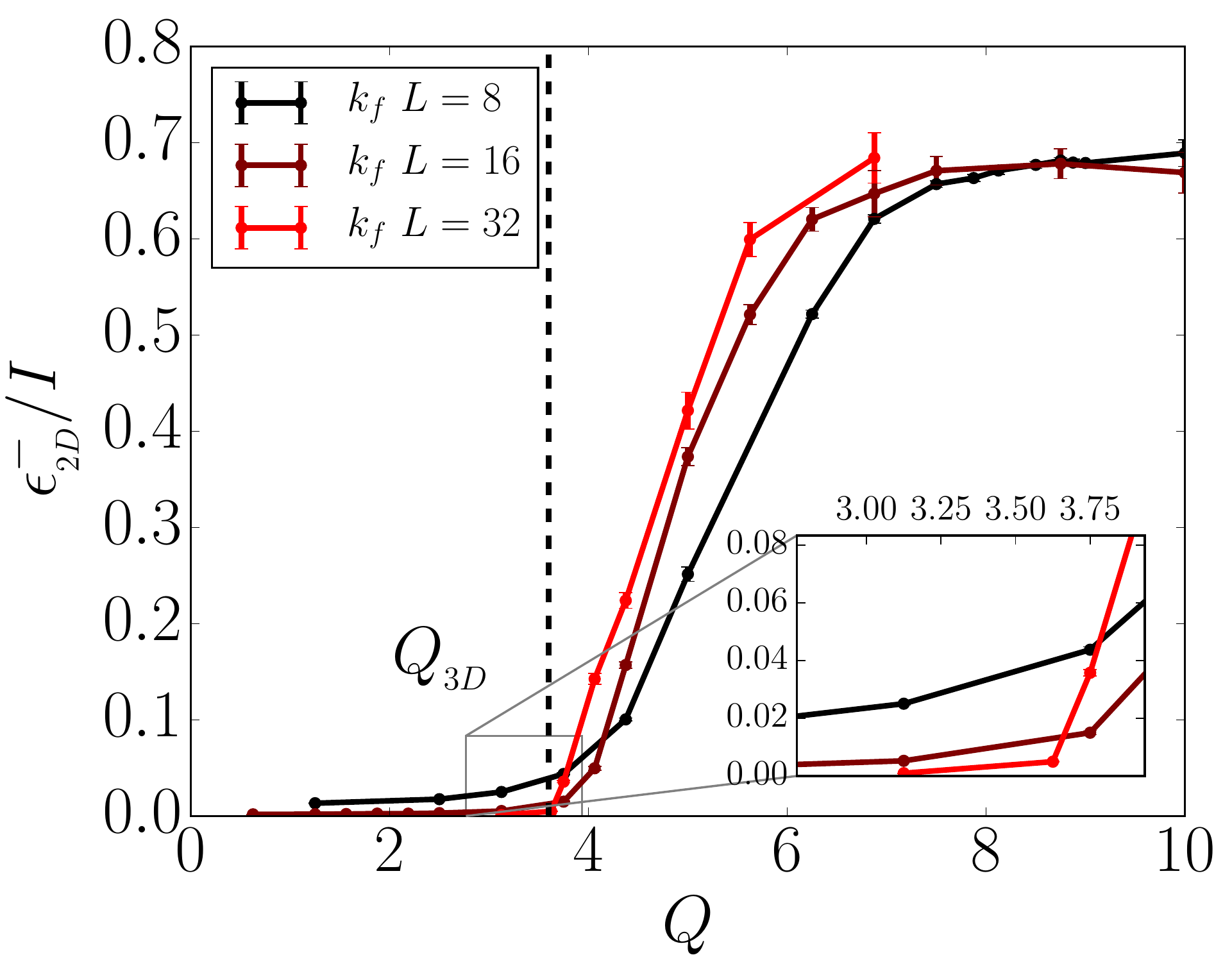}
		\includegraphics[width=0.48\textwidth]{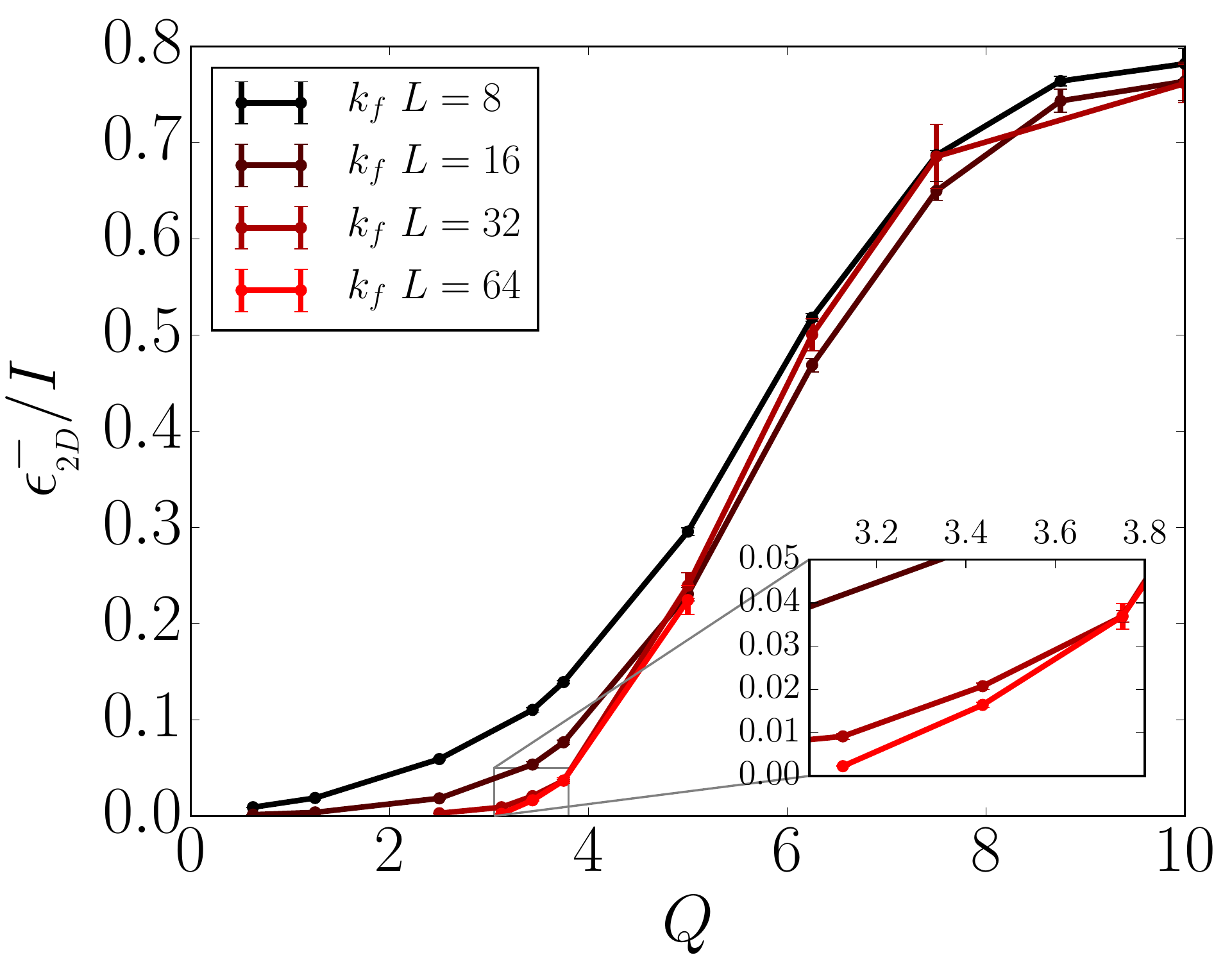}
		\caption{{\bf Left:} Runs from Case A. Measure of the inverse cascade for various runs as we increased the boxed size. We see that our runs converge to a sharp transition in the large box limit,
			confirming the existence of $\QTC$. 
			The dashed vertical line represents the approximate value of $\QTC$. {\bf Right:} Runs from Case B. A similar tendency is seen, although convergence is not yet reached.}
		\label{Q3D_Fig}
	\end{center}
\end{figure*}

We first investigate the first critical point $\QTC$ that marks the transition from a forward cascade to bidirectional cascade.
Figure \ref{Q3D_Fig} shows the dependence of the inverse cascade measured by the ratio of $\eom/I$ 
as a function of $Q$ for flows driven by the deterministic forcing \ref{det} 
(left) and by constant injection of energy forcing \ref{constI} (right). Since both cases are similar, we focus on the left subfigure, Case A.
The three lines correspond to three different box sizes $k_fL$ as marked in the legend.
For the smallest box size $k_fL=8$ the transition appears smooth with the presence of an inverse cascade at even the smallest 
values of $Q$ displayed. As the box size is increased the transition appears to become sharper, and at the largest 
box size the system seems to converge into a critical transition at the value $Q=\QTC\simeq 3.6$.
Close to the critical point the inverse cascade $\eom/I$ appears to scale linearly with the deviation from criticality 
$\eom \propto (Q-\QTC ) I$ for $Q>\QTC$.
Looking at Case B on the right of Fig. \ref{Q3D_Fig}, the transition seems to take longer to converge but the same tendency is seen.

\begin{figure}                                                                                            
  \begin{center}\vspace{0cm}                                                                              
    \includegraphics[width=0.45 \textwidth]{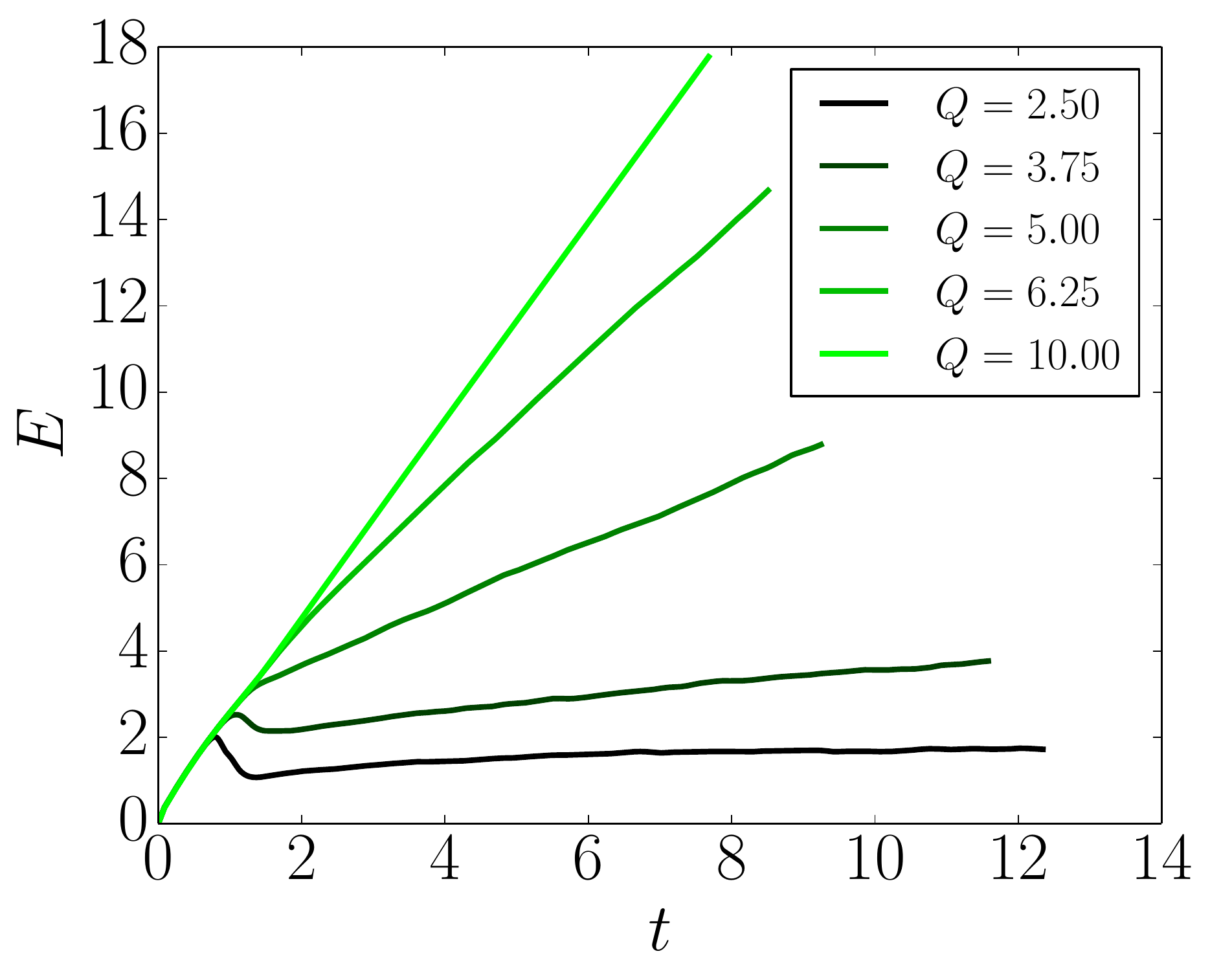}\vspace{0.5cm}                                        
    \includegraphics[width=0.45 \textwidth]{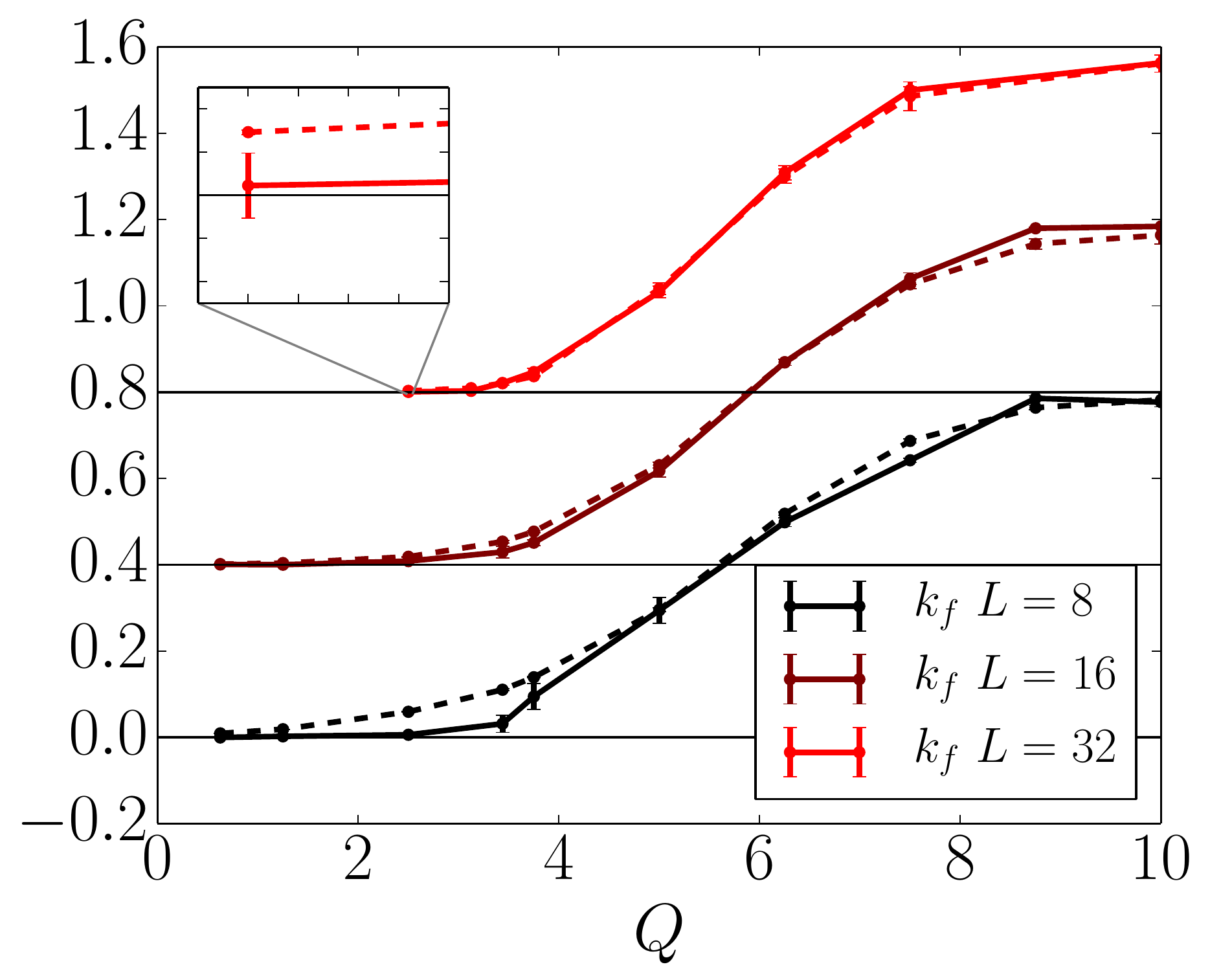}\vspace{0.5cm}                                 
	  \caption{{\bf Left:} The evolution of the energy for different values of $Q$ in the absence of large   
	  scale dissipation. {\bf Right:} The results of our constant energy injection runs, comparing the 
	  two methods of measuring the inverse cascade. The solid lines represent the slope measurement with no 
	  hypodissipation, whereas the dashed lines represent the steady-state long-time averaged measurements  
	  of runs with hypodissipation (Case B). The plots are shifted to emphasize differences. The horizontal black    
	  lines all correspond to $\eom = 0$ for the corresponding plot.}                                       
	\label{HypoVsSlope_Fig}                                                                                 
  \end{center}                                                                                            
\end{figure}

An alternative way to quantify the rate energy cascades towards the large scales is by measuring the rate of increase 
of energy. An inverse cascading system without large-scale dissipation and with a constant energy injection rate is expected
to lead to a linear increase of energy with time. This linear increase is expected after a short transient time where small scales reach a quasi-steady state
and before the largest scales of the system are exited and a condensate starts to form. 
The growth rate of the energy (i.e. the linear slope) is due to the inverse cascade and is equal to the inverse energy flux.
We tested this method of measurement by performing runs identical to those of Case B (using the forcing with constant energy injection given in Eq. \ref{constI}) 
but this time {\it without} a hypo-dissipation term, $\mu=0$.  
The energy evolution for the runs without hypo-dissipation are depicted in the left panel of Fig. \ref{HypoVsSlope_Fig}  
for different values of  $Q$, where a linear increase of energy can be seen. The slope of the linear growth of energy
for different values of $Q$ and $k_f L$ is calculated. These results are compared 
with the results obtained from steady state runs in the presence of hypo-dissipation
and are shown in the right panel of Fig. \ref{HypoVsSlope_Fig} for three different values of 
the box size. For the smallest value of $k_f L$ the two measurements differ quite a bit, with the steady state hypo-viscous simulations (dashed) which display a much more smooth behavior, 
and the $\mu = 0$ runs (solid) having significantly larger error.
As the box size is increased the two methods of
measuring the amplitude of the inverse cascade converge and the two curves overlap. 
Although measuring the inverse cascade by the slope of the energy versus time graph converges faster to 
the large box limit, the calculation using the hypo-dissipation term leads to much smaller 
error-bars, due to the long time averaging that is possible in this case. 
The smaller error is favorable when attempting to determine very small trends of inverse cascade, as is seen Fig. \ref{Q3D_Fig}.
%

\begin{figure}                                                                                                    
  \begin{center}\vspace{0cm}                                                                                      
    \includegraphics[width=0.49\textwidth]{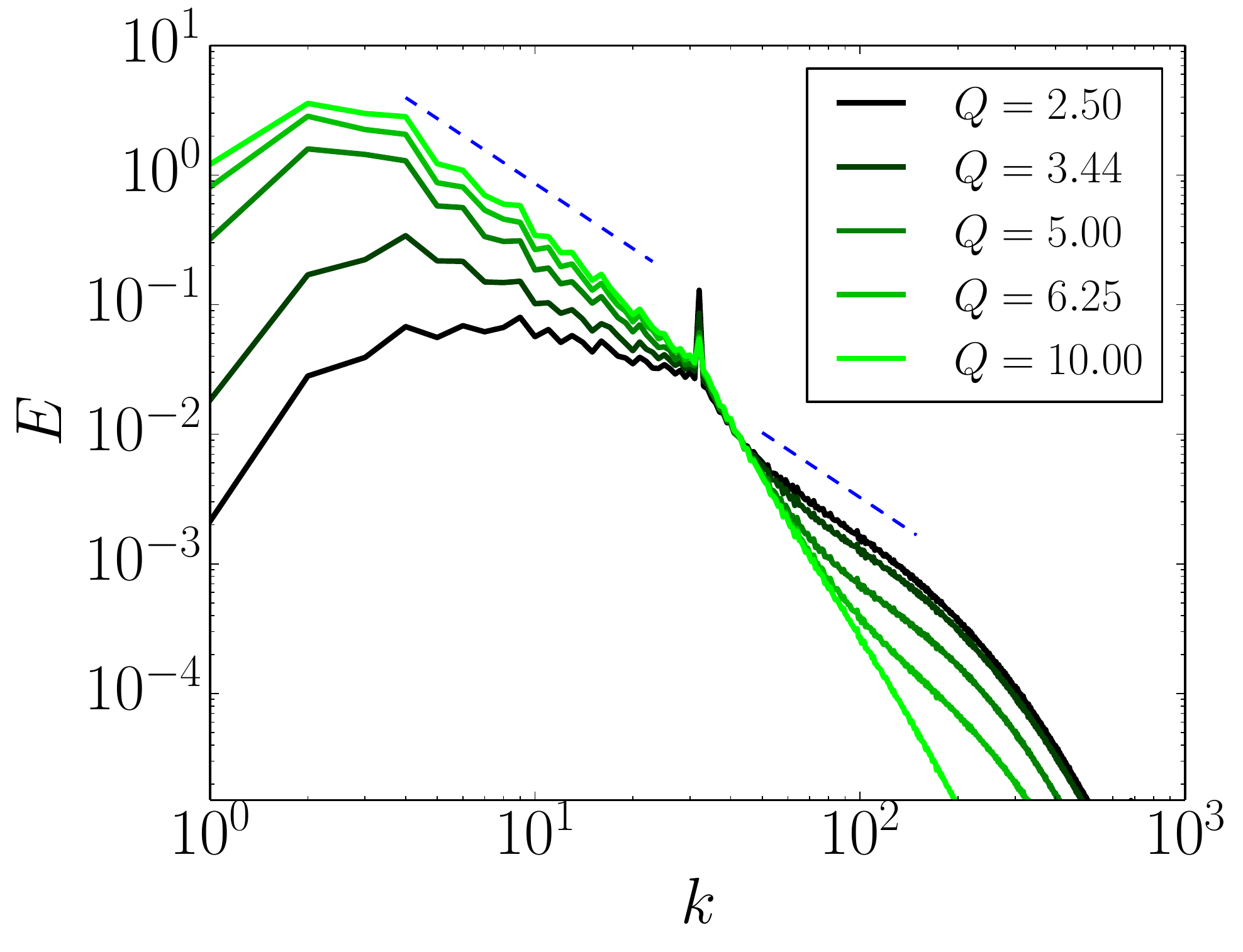}\vspace{0cm}                                        
    \includegraphics[width=0.49\textwidth]{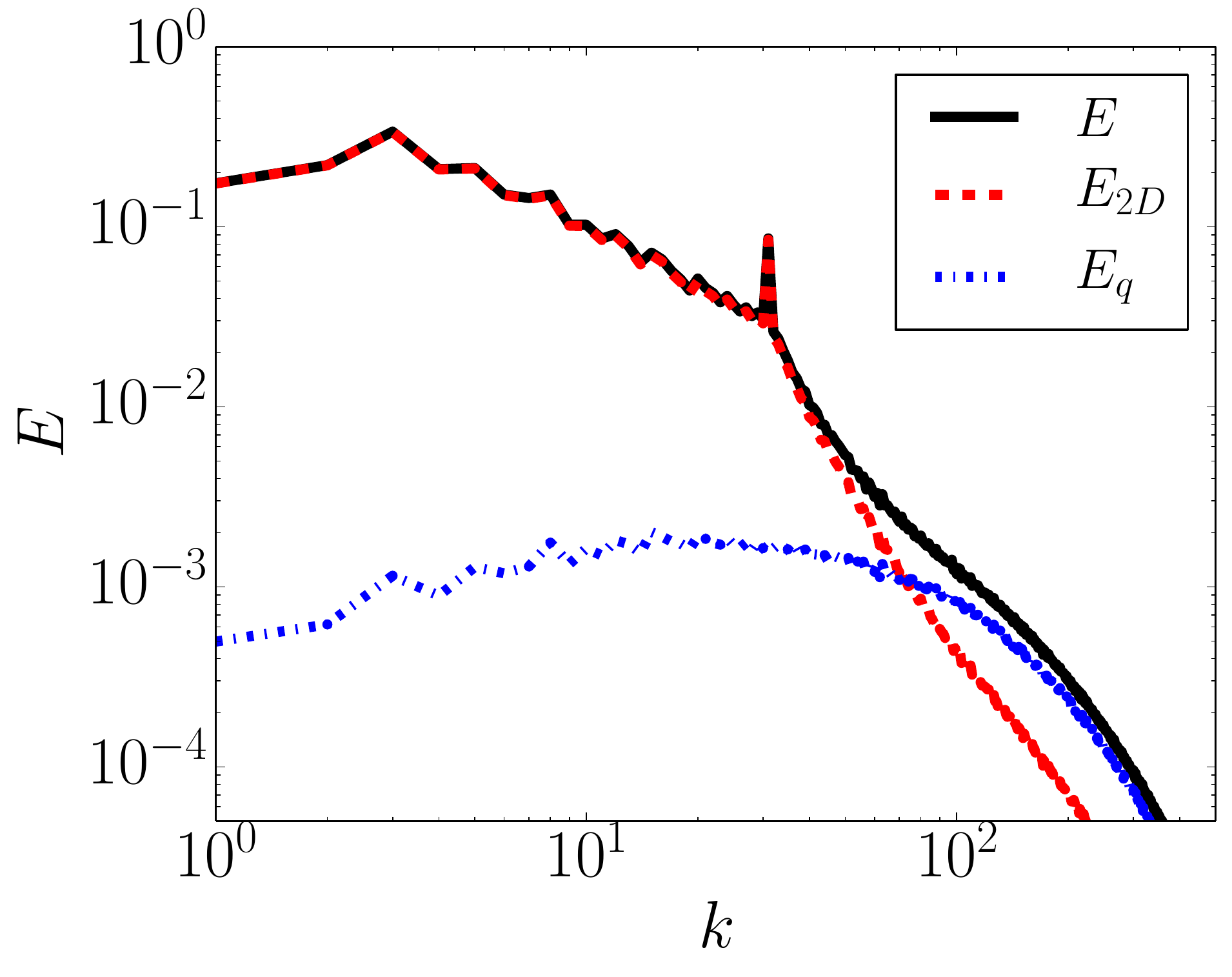}\vspace{0cm}                                        
    \caption{{\bf Left:} Spectra of the total energy for different values of $Q$ (averaged over many outputs).    
    The color of the curves become lighter and lighter as we have more and more inverse cascade.                  
    {\bf Right:} The total energy $E(k)$ (black), the two-dimensional energy $E_{2D}$                             
    (red), and the 3D energy $E_{q}$ (blue) for $Q=3.44$.}                                                    
	\label{SpecVsQD8S3Fig_Fig}                                                                                      
  \end{center}                                                                                                    
\end{figure}

We next look how the spectral distribution of energy is changing as $Q$ is varied by looking at the energy spectra $E(k), E_{2D}(k),$ and $E_{q}(k)$ 
defined as:
\begin{align}                                                                                        
E_{2D}(k) &= \frac{1}{2}\sum_{k<|{\bf k}|\le k+1}  |\tuo({\bf k})|^2, \\                               
E_{q} (k) &= \frac{1}{4}\sum_{k<|{\bf k}|\le k+1} |\tvq({\bf k})|^2,
%
\end{align}                                                                                          
where $\tuo({\bf k})$ and $\tvq({\bf k})$ stand for the Fourier mode amplitude of wavenumber 
$\bf k$ of the $\uo$ and $\vq$ fields respectively. 
The total energy spectrum $E(k) = E_{2D}(k) + E_{q}(k) $ is shown in the left panel of Fig. \ref{SpecVsQD8S3Fig_Fig}
for different values of $Q$ and for $k_fL=32$ .
These spectra were outputted during the run time of our simulations and then afterwards averaged together during the steady state period to get the final results. 
The smaller values of $Q$ are displayed with darker colors while the larger values of $Q$ are displayed with lighter colors.
Clearly as $Q$ is increased there is more energy in the large scales and less energy in the small scales.
For the smallest value of $Q$ the energy spectrum is decreasing for $k\ge k_f$.  This decrease of the energy spectrum is
compatible with a $k^{-5/3}$ power law scaling of a three-dimensional flow even though in our case only one mode with three-dimensional structure has been kept.
As we increase $Q$ we see the large scales gaining more and more energy until we reach the point where we now see a $k^{-5/3}$ spectrum at scales larger than $1/k_f$, 
as observed in 2D turbulence.
%
The right panel of Fig. \ref{SpecVsQD8S3Fig_Fig} compares the three spectra $E_{q}$ (blue, vertical-dashed line) and $E_{2D}$ (red, dashed line) for the intermediate value of  $Q=3.44>\QTC$.
At large wavenumbers the energy spectrum is dominated by $E_{q}$, while at small wavenumbers, which exhibit some inverse cascade, the spectrum is dominated by
$E_{2D}$.

The direction of the cascade is best described by 
the direct measurement of the transfer of energy among scales provided by the nonlinear terms in our equations.
More precisely, this flux of energy expresses the rate energy is transfered out of a given set of
wavenumbers due to the nonlinearities.
By performing a filtering of the 2D velocity field $\uo$ and the 3D velocity field $\uq$ in Fourier space
so that only the wavenumbers with modulus smaller than $k$ are kept, denoted by $\uo^{<k}$ and $\uq^{<k}$, one looks at only the structures of scales larger than $\ell = 2 \pi / k$. With these filtered velocity one can calculate the flux of energy, defined to be: 
\begin{align}
\Pi_{2D}(k) &= -\langle \uo^{<k} \cdot (\uo \cdot \nabla \uo)  \rangle, \\
\Pi_{q}(k) &= -\langle \uq^{<k} \cdot (\uo \cdot \nabla \uq)  \rangle, \\ 
\Pi_{T }(k) &= -\langle \uo^{<k} \cdot (\uq \cdot \nabla \uq)  \rangle - \langle \uq^{<k} \cdot (\uq \cdot \nabla \uo)  \rangle.
\end{align}
The first  flux $\Pi_{2D}$ expresses the rate energy is transfered from large-scale $\uo$ modes to small-scale $\uo$ modes by self-interaction.
The second flux $\Pi_{q}$ expresses the rate energy is transfered from large-scale $\uq$ modes to small-scale $\uq$ modes
through interactions with the 2D field $\uo$. Finally, the last flux $\Pi_{T }$ expresses the rate energy is transfered to the small scales
by a simultaneous exchange of energy from one field to the other. The total energy flux is given by
\beq
\Pi_{E}(k)=\Pi_{2D}(k)+\Pi_{q}(k)+\Pi_{T}(k).
\eeq
In the inertial range $\Pi_{E}(k)$ is constant and positive if the cascade is forward and
constant and negative if the cascade is inverse. In the case for which 
$\uq=0$
and enstrophy is conserved 
we can also define the flux of enstrophy as
 \beq
 \Pi_{\Omega}(k) = -\langle \nabla \times \uo^{<k} \cdot (\uo \cdot \nabla (\nabla \times \uo))  \rangle.
 \eeq
We note however that if $\uq$ is not exactly zero the enstrophy flux $ \Pi_{\Omega}(k) $ is not constant 
due to the enstrophy generation by the vorticity stretching term $\overline{\uq \cdot \nabla \uq}$.   
 
\begin{figure}
  \begin{center}\vspace{0cm}
    \includegraphics[width=0.48\textwidth]{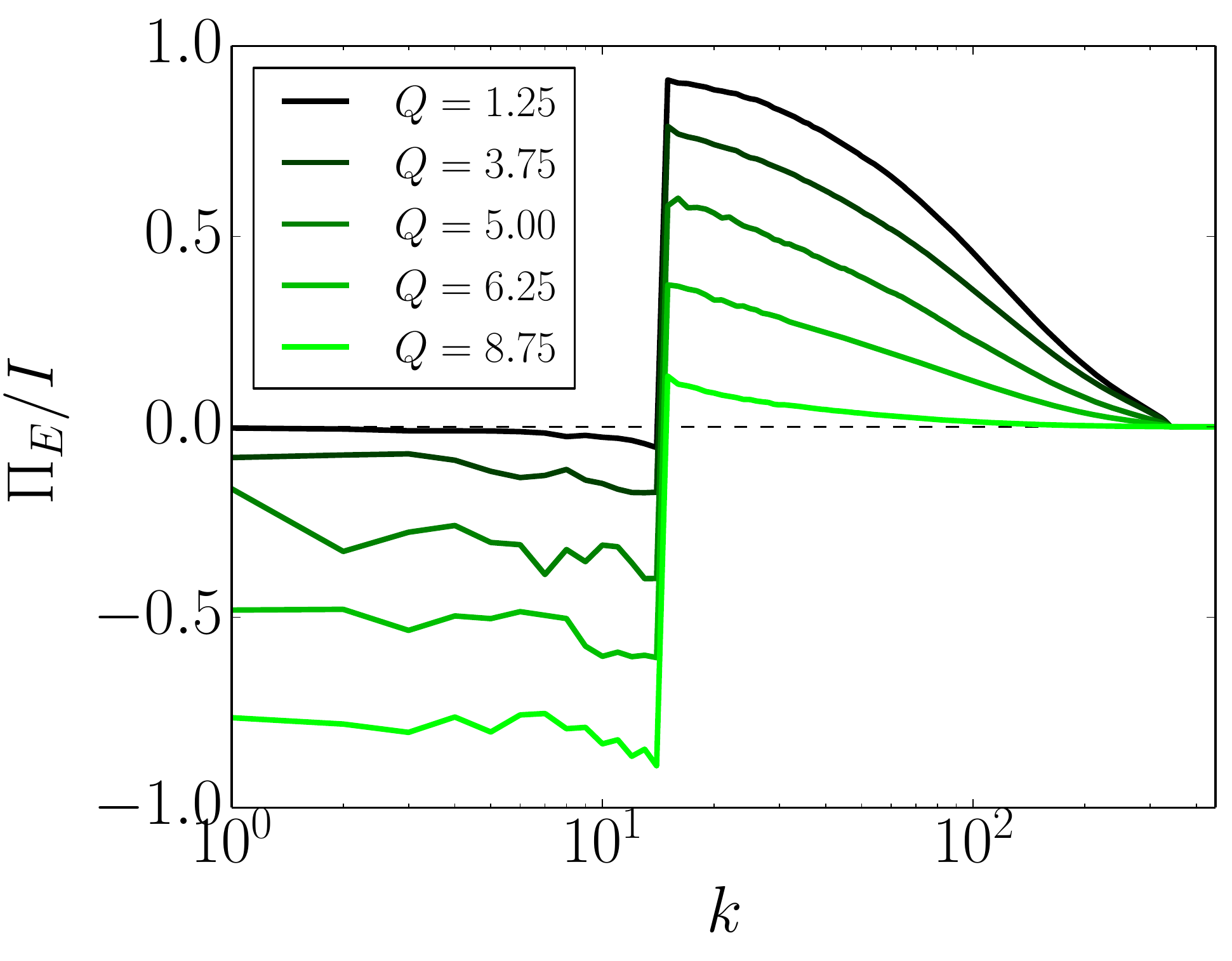}\vspace{0cm}\hspace{0.2cm}
    \includegraphics[width=0.48\textwidth]{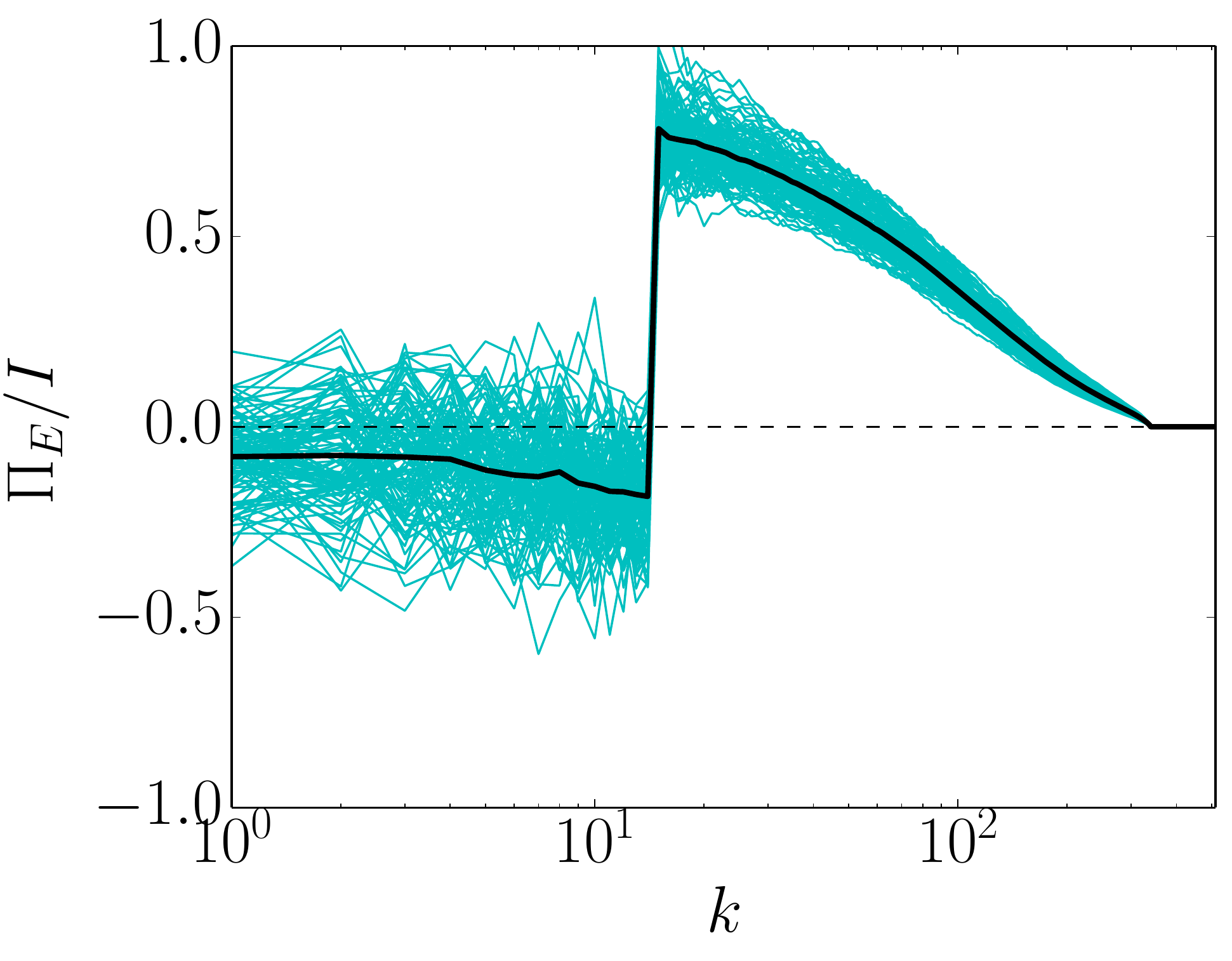}
	  \caption{{\bf Left:} The total energy flux $\Pi_E$ for different values of $Q$  (averaged over many outputs). The lighter the shade of green, the higher the value of $Q$. {\bf Right :} For the case of $Q = 3.75$: The cyan/light blue lines show the instantaneous values of the total energy flux; the solid black line is their average, corresponding to one of the curves on the left panel.}
	\label{FluxEvsQ}
  \end{center}
\end{figure}

Fig. \ref{FluxEvsQ} shows the time-averaged total energy flux $\Pi_E$ normalized by the energy injection rate $I$ for different values of $Q$. 
As before, dark lines correspond to small values of $Q$ while light lines correspond to larger values of $Q$. 
For the smallest value of $Q$ the flux $\Pi_{E}$ is almost zero for $k < k_f$ and positive for $k>k_f$ describing a unidirectional forward cascade.
As $Q$ is increased a bidirectional cascade appears with both positive and negative fluxes in either side of the forcing wavenumber
indicating that energy cascades to both large and small scales. At the largest value of $Q$ almost all energy cascades inversely
demonstrated by the negative values of the flux for $k<k_f$ and almost zero values for $k>k_f$.
The time averaged flux shows the predicted constant-in-$k$ behavior for small $k$ while it is affected by viscosity in the large scales
and thus decays with $k$. Instantaneous fluxes however are strongly fluctuating. This is displayed in the right panel of Fig. \ref{FluxEvsQ},
where the time-averaged value of the 
flux (thick black line) is displayed along with numerous 
instantaneous fluxes (cyan lines) for the case of $Q= 3.75$. The role of the fluctuations becomes particular important close to the transition
where the amplitude of the fluctuations becomes much larger than the mean value. 

\begin{figure}
  \begin{center}\vspace{0cm}
    \includegraphics[width=0.45\textwidth]{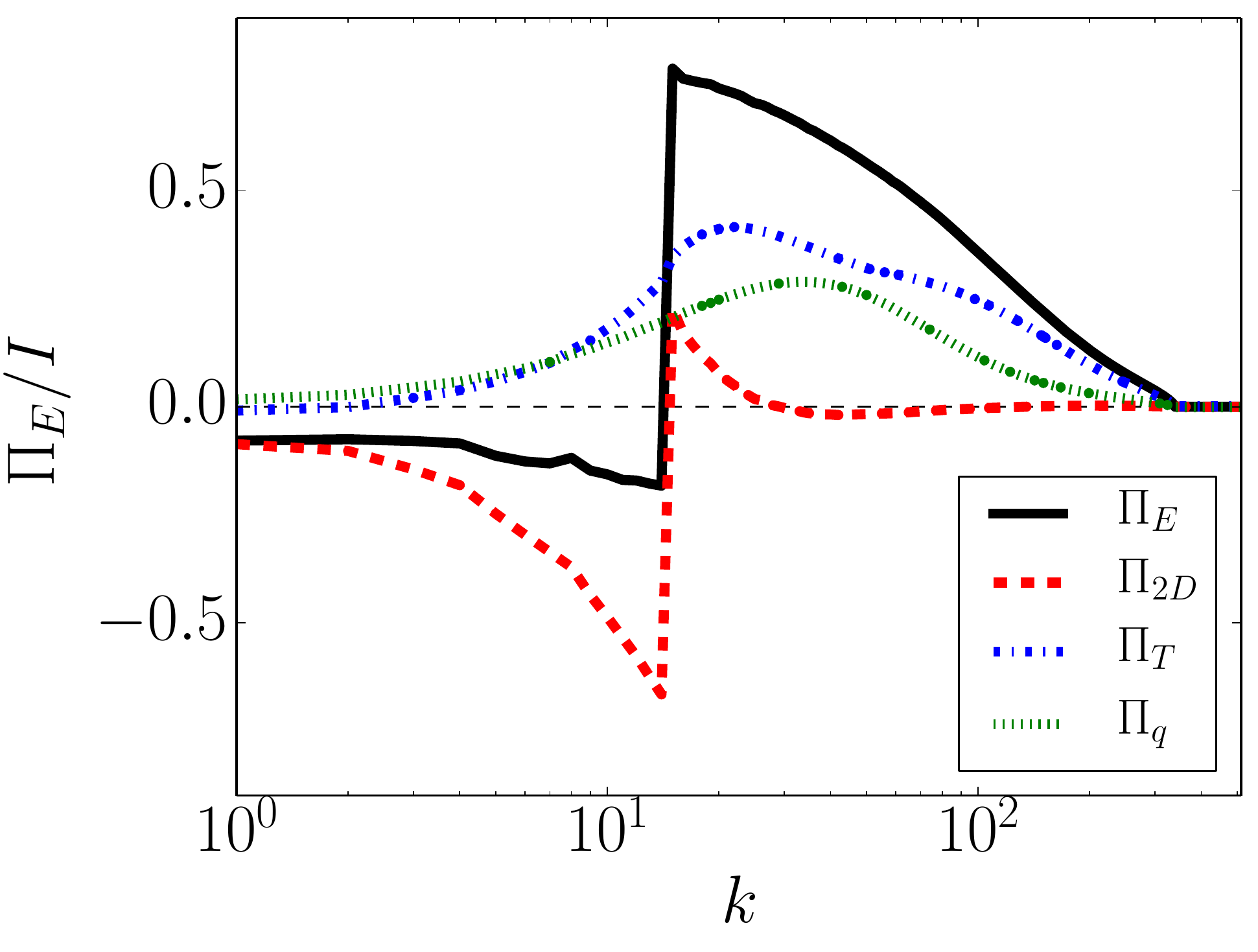}\hspace{0.2cm}
    \includegraphics[width=0.45\textwidth]{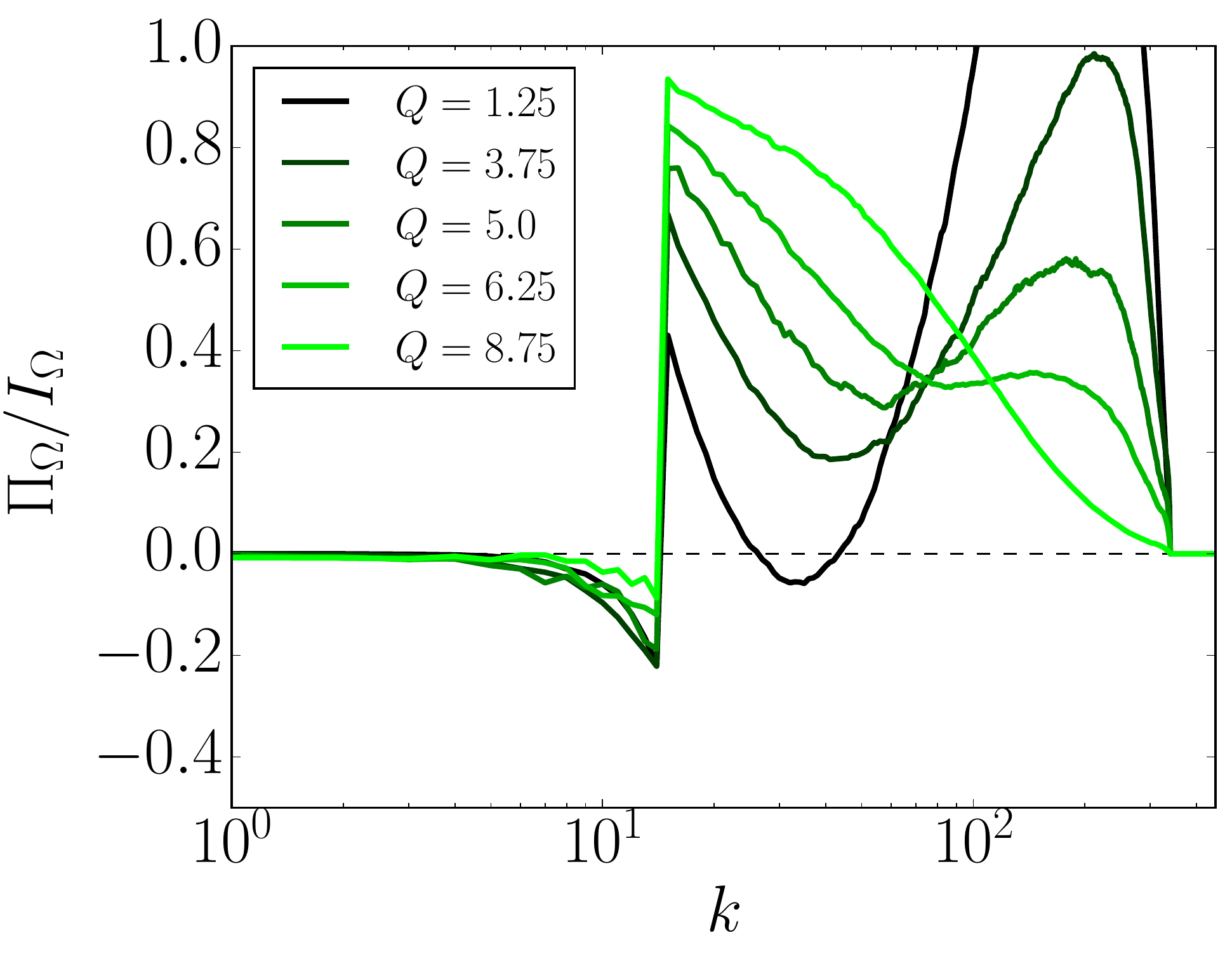}\vspace{0cm}
	  \caption{{\bf Left:} For the case of $Q = 3.75$: The total energy flux is seen in the thick black line, and the contributions from each nonlinear term is seen in the other colored lines. Notice that $\Pi_{2D}$ is the only one with a negative contribution.
	  	One should also note that the green dotted line, corresponding to the term $\Pi_{q}$, is always positive for all $Q$ examined but that the blue dash-dot line, corresponding to $\Pi_{_T}$, can be both negative and positive depending on $Q$ and $k$. 
	  	{\bf Right:} The enstrophy flux $\Pi_\Omega$ for different values of $Q$  (averaged over many outputs)
	  	normalized by the enstrophy injection rate $I_\Omega=\langle \nabla \times \uo \cdot \nabla \times {\bf F} \rangle_{_T}$. 
	  	The color of the curves become lighter and lighter as the value of $Q$ increases.}
	\label{FluxComponents}
  \end{center}
\end{figure}

To demonstrate the role of each field in the cascade of energy we show, in 
the left panel of Fig. \ref{FluxComponents}, the decomposition of the total energy flux $\Pi_E$ into the three 
components $\Pi_{2D},\Pi_{q}$, and $\Pi_{T}$ for the value of $Q=3.75$. 
The flux $\Pi_{2D}$ is negative for all values of $k<k_f$ while it is positive but small for $k>k_f$.
The remaining fluxes $\Pi_{q}$ and $\Pi_{T}$ are positive for all $k$. 
Thus the inverse cascade is driven by the  $\Pi_{2D}$ term 
%
%
whereas the forward cascade is driven mostly by $\Pi_{q}$ and $\Pi_{T}$.
It is worth noting that neither of these partial fluxes is constant in the inertial range.
Furthermore $\Pi_{q}$ and $\Pi_{T}$ are positive and finite for $k<k_f$ implying that
part of the energy transfered to the large scales by the $\uo$ field is brought back to the small 
scales by interactions with the $\uq$ field.

The right panel of Fig. \ref{FluxComponents} shows the flux of enstrophy $\Pi_{\Omega}$ for different values of $Q$. 
As discussed in the previous section enstrophy is only conserved by the nonlinearities  when $\uq=0$.
For large values of $Q$, for which $\uq$ is small, we expect that enstrophy will be quasi-conserved and its flux will be positive and slowly varying.
This is indeed observed in this figure for $Q=8.75$, where the enstrophy flux is slowly decreasing due to viscous effects.  
For smaller values of $Q$, however, the enstrophy flux becomes non-monotonic with a sharp increase close to the 
dissipation scales. This is due to the generation of enstrophy by the $\uq$ field that leads to an excess of enstrophy 
that is transported to the small scales. 

\begin{figure}                                                                                   
    \centering                                                                                   
    \begin{subfigure}[b]{0.35\textwidth}                                                         
        \includegraphics[width=\textwidth]{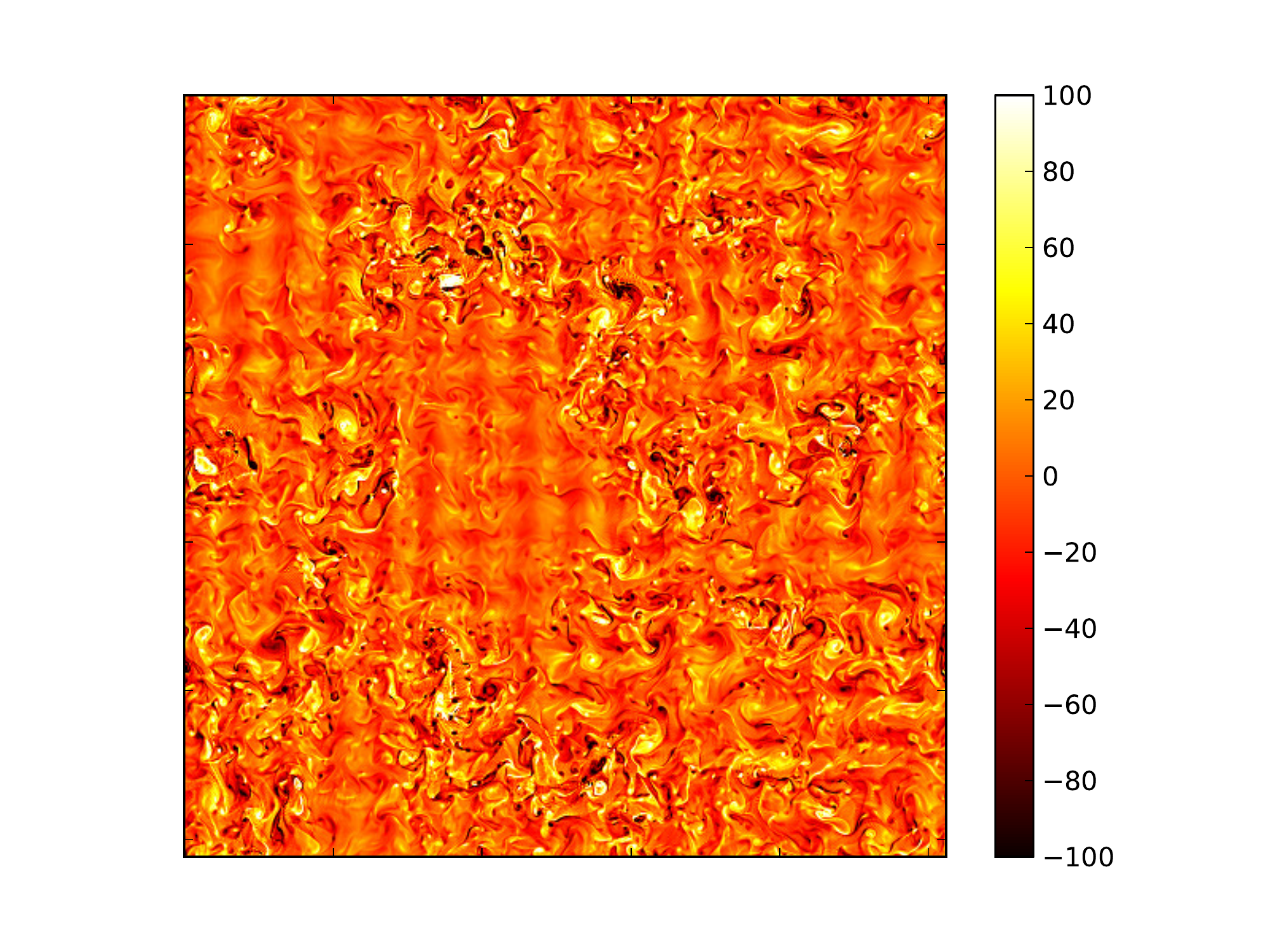}                                
    \end{subfigure}\hspace{-0.7cm}                                                               
    \begin{subfigure}[b]{0.35\textwidth}                                                         
        \includegraphics[width=\textwidth]{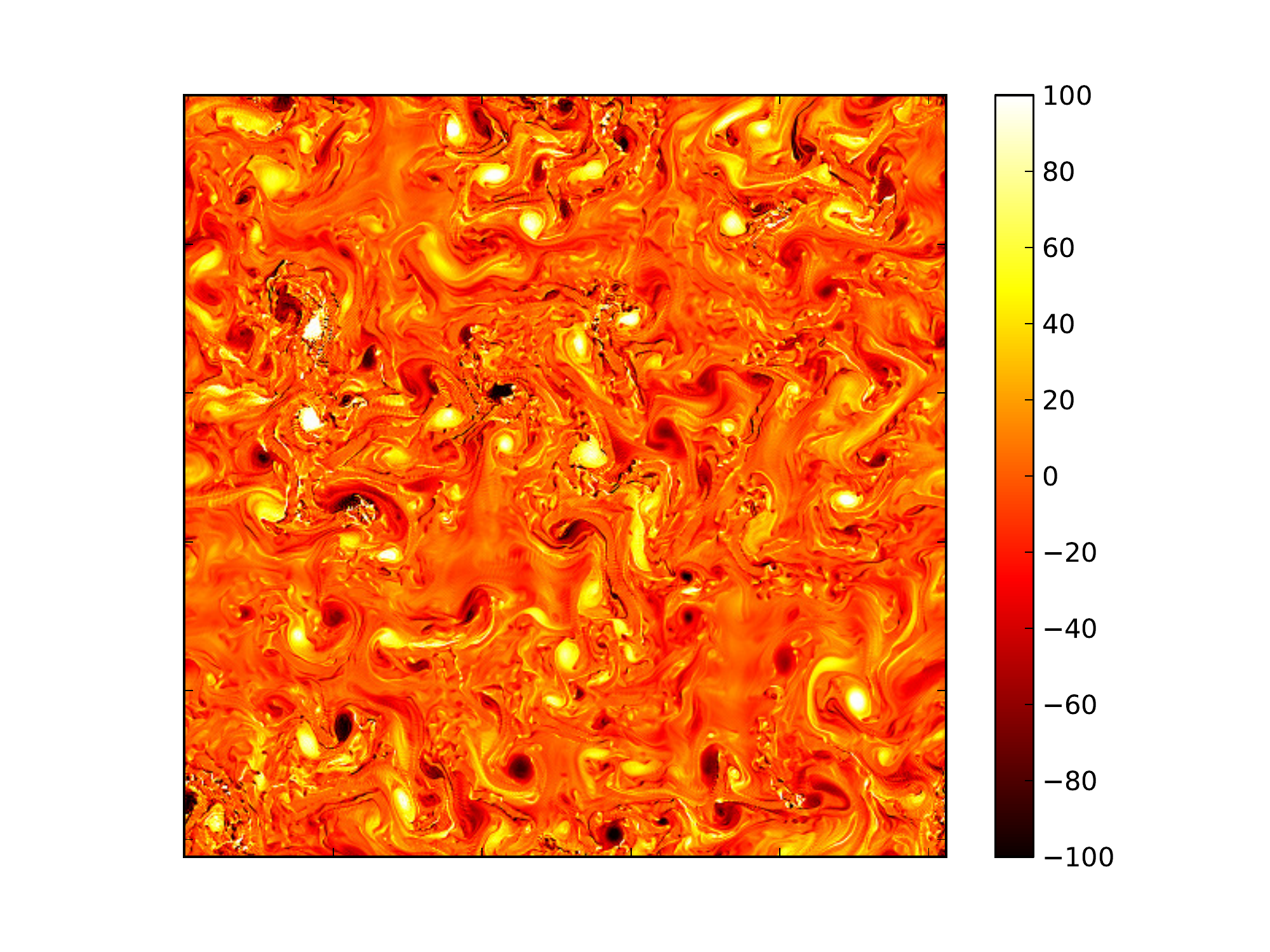}                              
    \end{subfigure}\hspace{-0.7cm}                                                               
    \begin{subfigure}[b]{0.35\textwidth}                                                         
        \includegraphics[width=\textwidth]{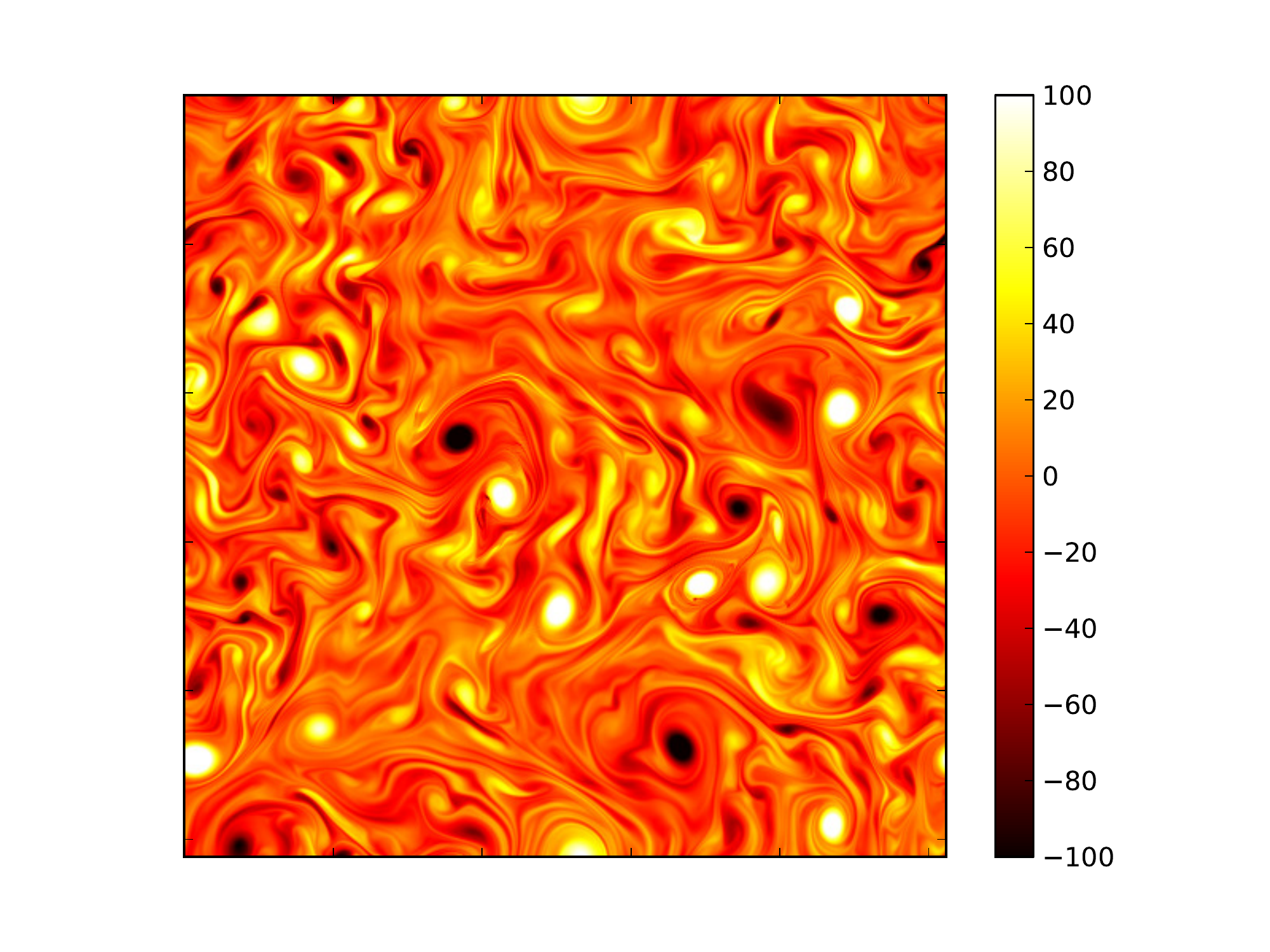}                               
    \end{subfigure}                                                                              
    \begin{subfigure}[b]{0.35\textwidth}                                                         
        \includegraphics[width=\textwidth]{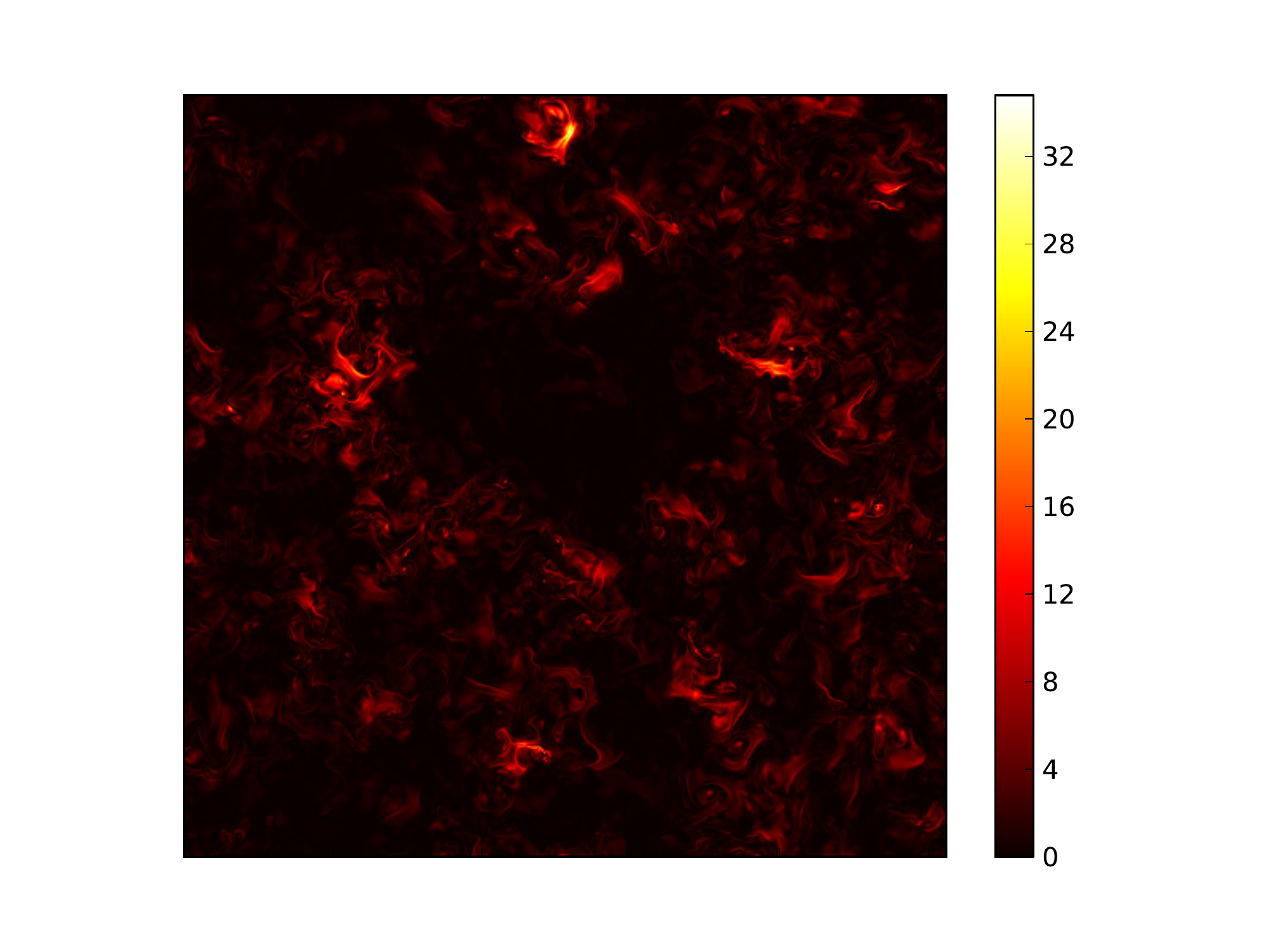}                             
        \caption{$Q = 1.25$}                                                                     
    \end{subfigure}\hspace{-0.7cm}                                                               
    \begin{subfigure}[b]{0.35\textwidth}                                                         
        \includegraphics[width=\textwidth]{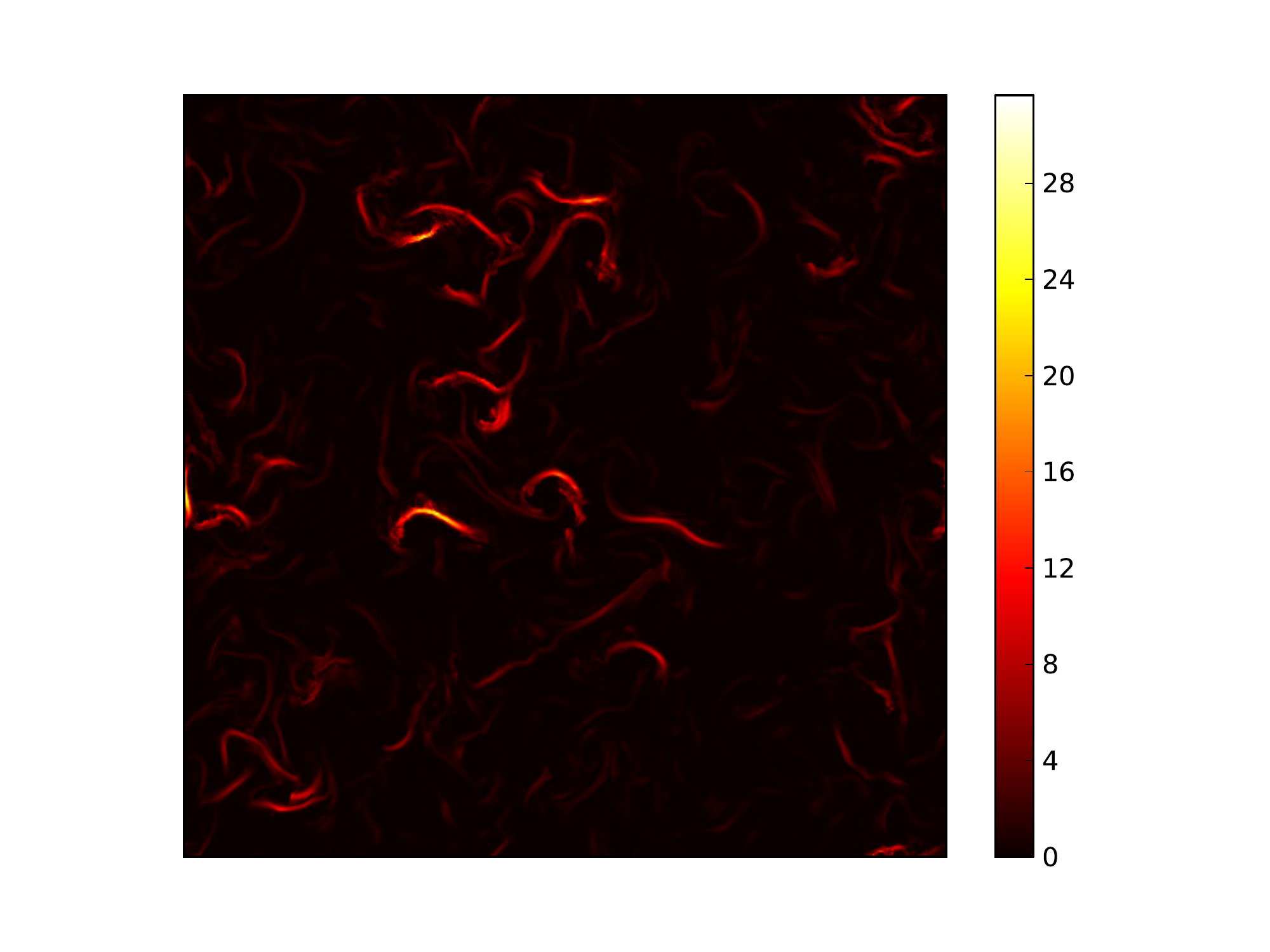}                           
        \caption{$Q = 3.75$}                                                                     
    \end{subfigure}\hspace{-0.7cm}                                                               
    \begin{subfigure}[b]{0.35\textwidth}                                                         
        \includegraphics[width=\textwidth]{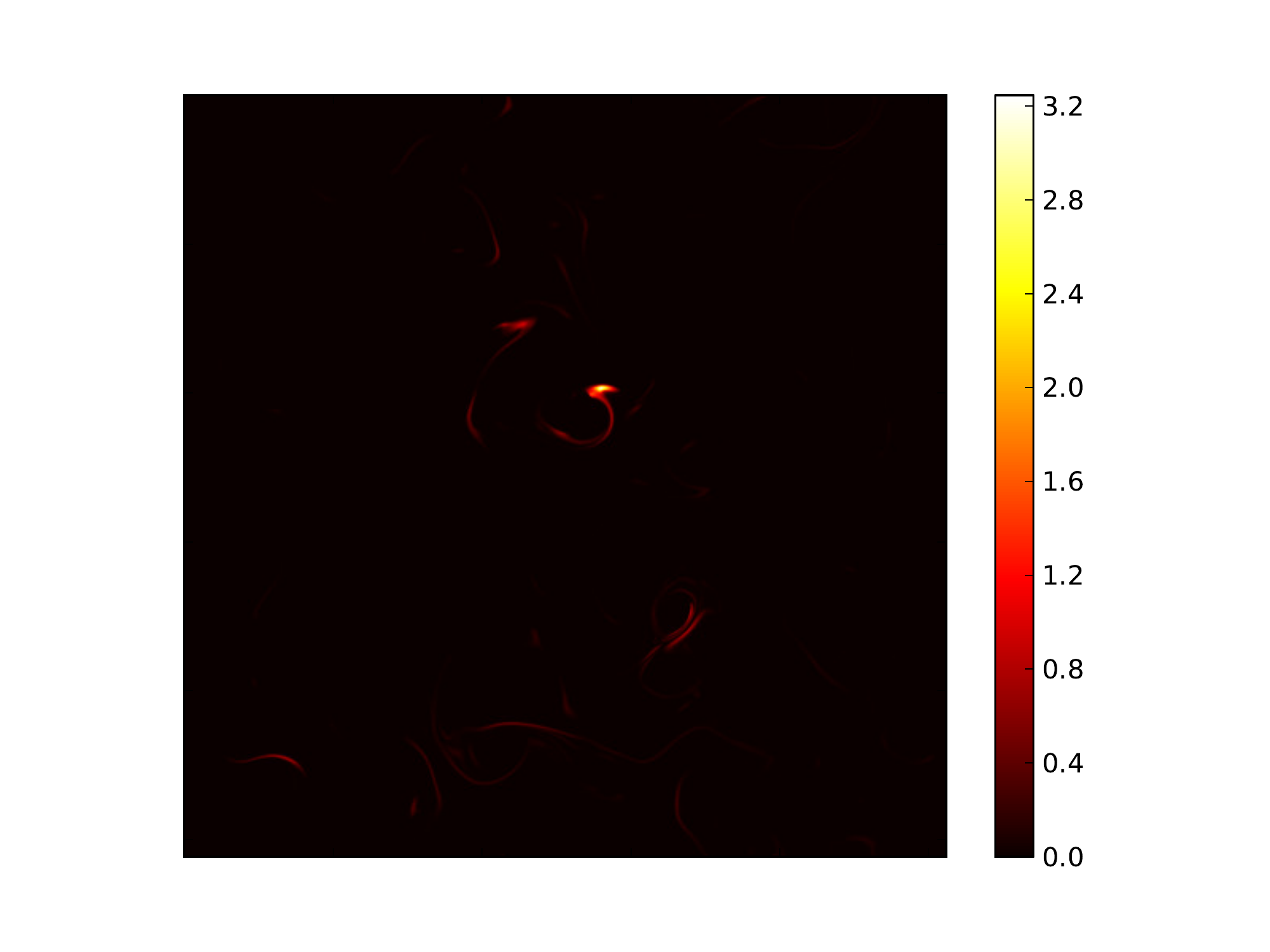}                            
        \caption{$Q = 8.75$}                                                                     
    \end{subfigure}                                                                              
    \caption{ The two-dimensional vorticity $\nabla \times \uo$ (top) and 3D energy density      
    $\mathcal{E}_{q}$ (bottom), for three values of $Q$. The box size $k_f L = 16$ is fixed      
    for all cases.                                                                               
    	The grid pattern which is seen in the top left image is due to the forcing.                
    	} \label{prettypix_3DEn}                                                                   
\end{figure}                                                                                     

The results so far have demonstrated the presence of a bidirectional cascade in Fourier space.
Further insight is obtained by looking at the resulting structures in physical space.
Fig. \ref{prettypix_3DEn} shows, in the top panels, the two-dimensional vorticity $\nabla \times \uo$ for three values of $Q$.
The bottom panels show the energy density 
%
$\mathcal{E}_{q} \equiv |\vq|^2 / 4$
of the 3D field at the same instant of time as the vorticity for the same values of $Q$.
The vorticity snapshots show that, as $Q$ is varied from small to large values, the structures change from mainly small-scale eddies 
to eddies larger than the forcing scale $k_f L = 16$,
further demonstrating the growth of the inverse cascade as $Q$ is increased.  
For the smallest value of $Q$ the structures are small compared to the forcing scale, and one can notice by the sharp changes from very bright to dark that there are also large gradients of vorticity.
For the largest value of $Q$ the flow has very similar structure to that obtained in pure 2D simulations.  
Finally, the intermediate value of $Q$ appears to have characteristics of both extreme cases but concentrated 
in different regions of space. Some regions in space resemble the 
small-scale-eddies and large-gradient
case of small $Q$ while other 
regions have the typical 2D eddy structures.
The structures in the 3D energy density snapshots, shown in the bottom panels of Fig. 
\ref{prettypix_3DEn} are filamentary and are well correlated with the areas of high strain in the 2D field. Furthermore, the density 
of these areas of high $\mathcal{E}_{q}$ 
appear to decrease with $Q$.
Combining these observations,
one notices that, for small values of $Q$, the intensity of vorticity is not uniform throughout the domain -- some regions appear to be 
`3D-active' and some regions to be `3D-quiet'.
As $Q$ is increased the `3D-active' regions appear to become less space filling. 
A similar behavior has been observed in the transition from forward to inverse cascade in 2D magneto-hydrodynamic flows,
where the role of $\uq$ was played by the magnetic field \citep{us1,us2}.  
This behavior, as we will discuss next, has direct consequences for the behavior of the flow close to the second critical point.

\section{Transition from a bidirectional to an inverse cascade}\label{sec:Q2D}   

We now focus on large values of $Q$ for which the flow transitions from a bidirectional cascade to an inverse unidirectional cascade.
As shown in the previous section the terms that involve the three-dimensional field $\uq$ are responsible for driving the forward cascade. 
In their absence we recover the 2D Navier-Stokes that leads to an inverse cascade with all energy dissipated at large scales in the large $Re$ limit.
Thus in order to transition to a unidirectional inverse cascade the flow needs to become purely 2D with $\uq=0$. This is indeed possible
since a purely 2D flow is always a solution of the 3D Navier-Stokes equations, Eq. \ref{NS}. 
Therefore if the initial data are exactly 2D the flow will remain 2D for all times cascading energy inversely. 
However these solutions can be unstable and small perturbations can grow exponentially driving the flow away from the 2D behavior
and thus alter the direction of cascade. This brings us to the intuitive understanding of the second critical point. 
The bidirectional cascade will transition to a unidirectional inverse cascade when the layer thickness is so  small ($Q$ is large enough) 
that all 3D perturbations are damped and decay exponentially in time, and thus the 2D solution is an attractor of the system. 
This is expected to occur when the thickness of the box $H_{_{2D}}$ is such that the viscous dumping rate $\nu/H_{_{2D}}^2$
due to the vertical variation alone is similar to the shear rate $u_f k_f$ that drives the 3D instability.
This argument $ \nu/H_{_{2D}}^2 \sim u_f k_f$ implies that the second critical point $\QDC$ will satisfy:
\begin{equation}\label{intuition_Q2D}
\QDC = \pi/(H_{_{2D}}k_f) \propto \sqrt{Re}.
\end{equation}

A more precise estimate for the location of $\QDC$ can be obtained by looking at the energy evolution of the 3D component of the flow
that reads
\begin{equation}\label{hami}
\frac{1}{2}\frac{d  }{dt} \la |\uq|^2 \ra  = - \la  \uq \cdot(\nabla \uo )\cdot \uq \ra  - \nu \la |\nabla \uq|^2 \ra  \equiv \mathcal H.
\end{equation}
Note that the equation above is also valid for the full Navier-Stokes equations, not just our model, with $\uq=\bf u -\uo$. 
The 2D solutions will be globally stable if the functional ${\mathcal H}$ 
is negative definite
in which case all 3D perturbations will decay independent of the initial conditions.
We can prove that this is the case for a given range of $Q$ using the following rigorous inequalities.
Using H\"older's inequality the vortex stretching term $\la  \uq \cdot(\nabla \uo )\cdot \uq \ra$ is bounded by  
\beq
  | \la  \uq \cdot(\nabla \uo )\cdot \uq \ra | \le \| \nabla \uo \|_\infty \la |\uq|^2 \ra
\eeq
where $\| \nabla \uo \|_\infty$ stands for the $L_\infty$ norm of the gradiends of $\uo$.
Furthermore Poincare's inequality gives, for the dissipation term, that $\la |\nabla \uq|^2 \ra \ge \nu q^2\la |\uq|^2 \ra $.
These two results lead to the negativity of $\mathcal H $ if $ Q^2 = k_f^2 q^2 \ge  \| \nabla \uo \|_\infty / k_f^2 \nu$.   
Provided that $\| \nabla \uo \|_\infty$ is bounded, this result guarantees that, beyond some value of $Q$, the flow becomes exactly 2D. 
However deriving an upper bound on $\| \nabla \uo \|_\infty$ in terms of the control parameters of the system requires some further analysis.
This has been achieved in \cite{gallet2015exact} where the two-dimensionalisation of a flow due to the
presence of a external strong magnetic field $B_0$ was examined. 
Their results extend directly to our case by setting $B_0=0$.
They were able to show, using properties of the steady state 2D Navier-Stokes equations \citep{alexakis2006energy}, that
the 3D energy will decay, in the long-time limit, if $Q$  satisfies:
\begin{equation}\label{Qcri_nonlin}
 Q^2 \geq  Re^3 \left(\frac{c_1}{Re} + c_2 \right),
\end{equation}
where each $c_i$ is a positive dimensionless number that depends on forcing shape and other non-universal properties of the flow. 
Note that $Q$ does not scale as we predicted with respect to $Re$.
This is because this bound is very conservative and doesn't capture all the physics at the transition point. 
However, despite this loose scaling, this results guarantees that the transition from a bidirectional to a purely inverse cascade will be through a critical point
because it guarantees that there is a value of $Q$ above which $\eqp$ is exactly zero.
 
Another less conservative bound which was derived in \cite{gallet2015exact} is based on the linear stability analysis of 2D flows to 3D perturbations. 
Linear stability guaranties that the 2D solutions are locally stable but does not exclude the possibility that a locally attracting 3D solution also exist for the same parameters.
Their analysis shows that 2D solutions (for harmonic forcing and in the absence of hypo-dissipation term)  are linearly stable provided that
\begin{equation}\label{lambdacase2}
Q^2 \geq  \ Re \left(c_3 + c_4 \ln|Re| + c_5 \ln\left|\frac{k_f}{2 \pi}\right|\right).
\end{equation}
Up to logarithmic corrections this result follows the scaling $\QDC \sim Re^{1/2}$, which is what our scale analysis predicts.

\begin{figure}                                                                                                            
  \begin{center}\vspace{0cm}                                                                                              
    \includegraphics[width=0.6\textwidth]{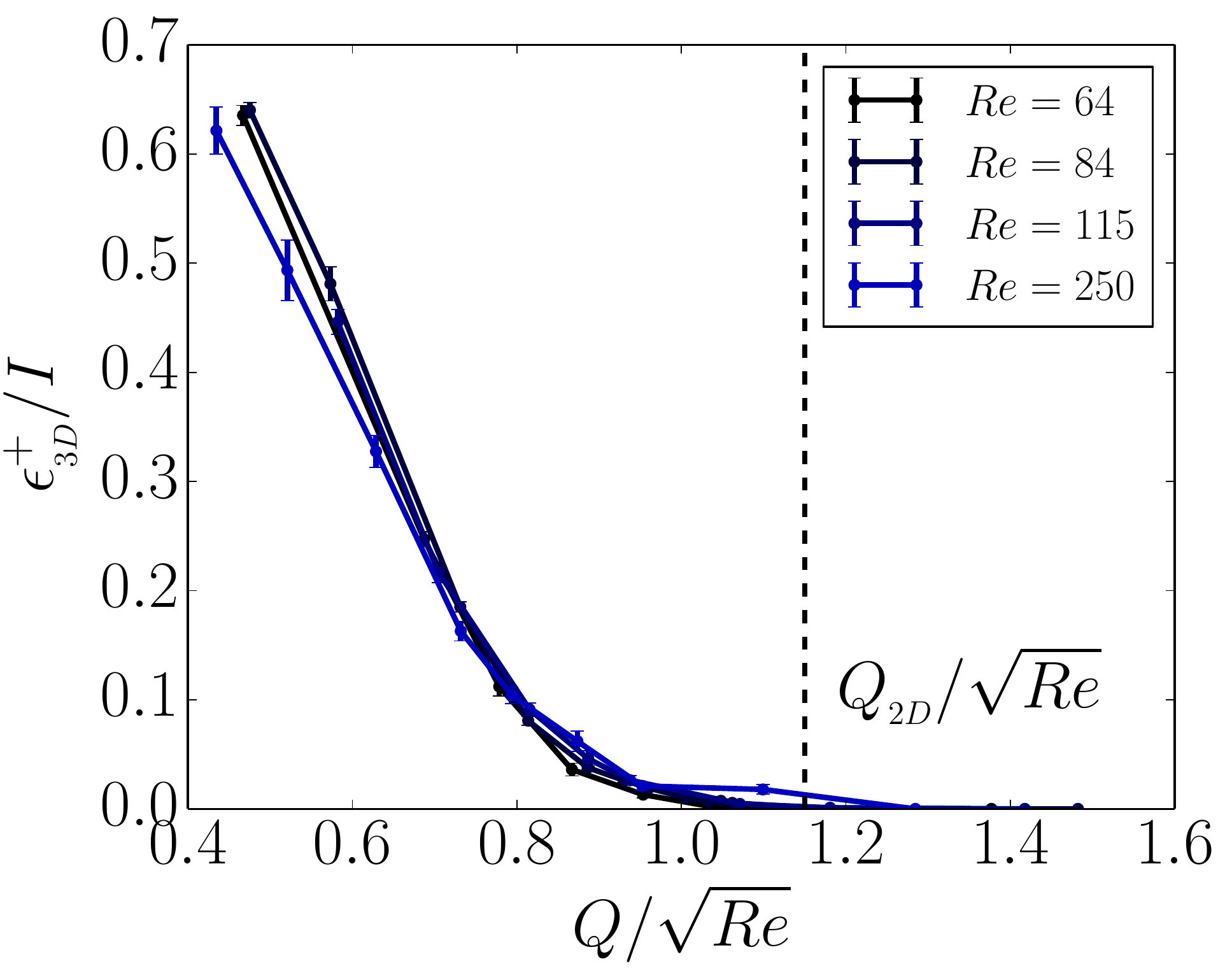}\vspace{0cm}                                                       
	  \caption{Run set C. $\eqp$ normalized by the energy injection rate                                                    
	  for various runs as we increased the Reynolds number. Rescaling $Q$ by our scale analysis prediction                  
	  collapses the curves onto one. The dashed vertical line represents the                                                
	  approximate value of $\QDC$, after which all runs have decaying 3D energy.                                    }       
	\label{Q2D_Fig}                                                                                                         
  \end{center}                                                                                                            
\end{figure}

This scaling is clearly demonstrated by our numerical simulations in Fig. \ref{Q2D_Fig} that shows
the energy dissipation rate $\eqp$ as a function of the rescaled $Q/\sqrt{Re}$. 
This rescaling of $Q$ collapses the curves together. The vertical dashed line marks the critical value $\QDC$
and represents the point after which 
the amplitude of the 3D field decays exponentially. 
In the absence of $\uq$ we have a 2D solution -- $\eqp$ is zero  
and most of the energy is dissipated in the large scales. 
Note that for a given value of $Q$ the exact 
two-dimensionalization occurs for the values of $Re$ in the range $Re \le Q^2$. Since $Re$ is bounded from above 
there is always some viscous dissipation $\eop$.
It is also worth pointing out that the dependence of $\eqp$ on the deviation from criticality differs 
from that of $\eom$ close to the first critical point $\QTC$ that was linear.
Close to $\QDC$ the energy dissipation $\eqp$ follows the relation
\begin{equation}
\eqp \propto (\QDC -Q )^\beta I, \quad \text{ where } \beta > 1  \, \text{ and } Q\le\QDC .
\label{sqr}
\end{equation}
The exponent $\beta$ was found to be close to $\beta\simeq 2$.  
To understand the origin of this exponent we need to look at the temporal and spatial form of the unstable field $\uq$. 

\begin{figure}                                                             
    \centering                                                             
    \begin{subfigure}[b]{0.49\textwidth}                                   
        \includegraphics[width=\textwidth]{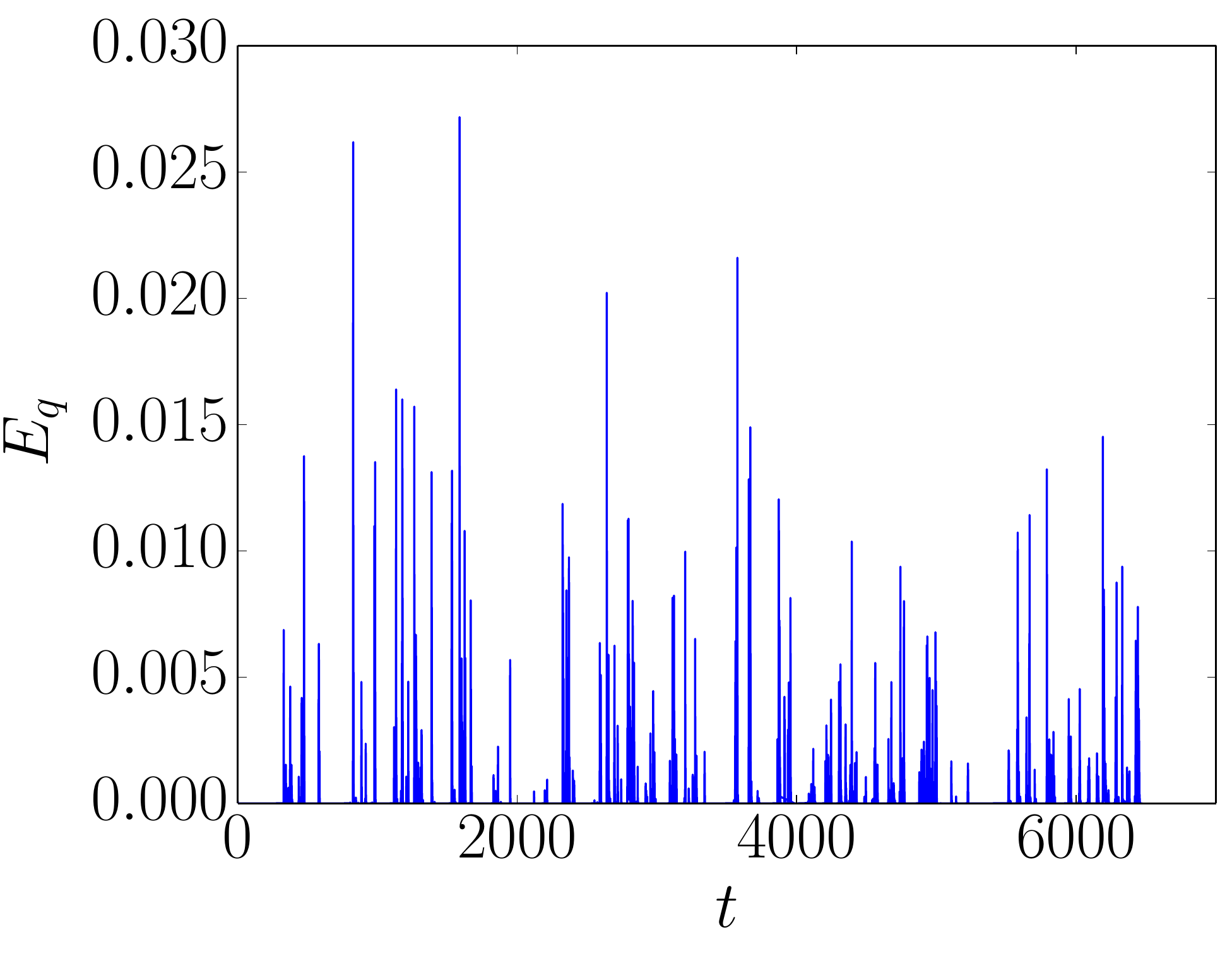}             
    \end{subfigure}                                                        
    \begin{subfigure}[b]{0.49\textwidth}                                   
        \includegraphics[width=\textwidth]{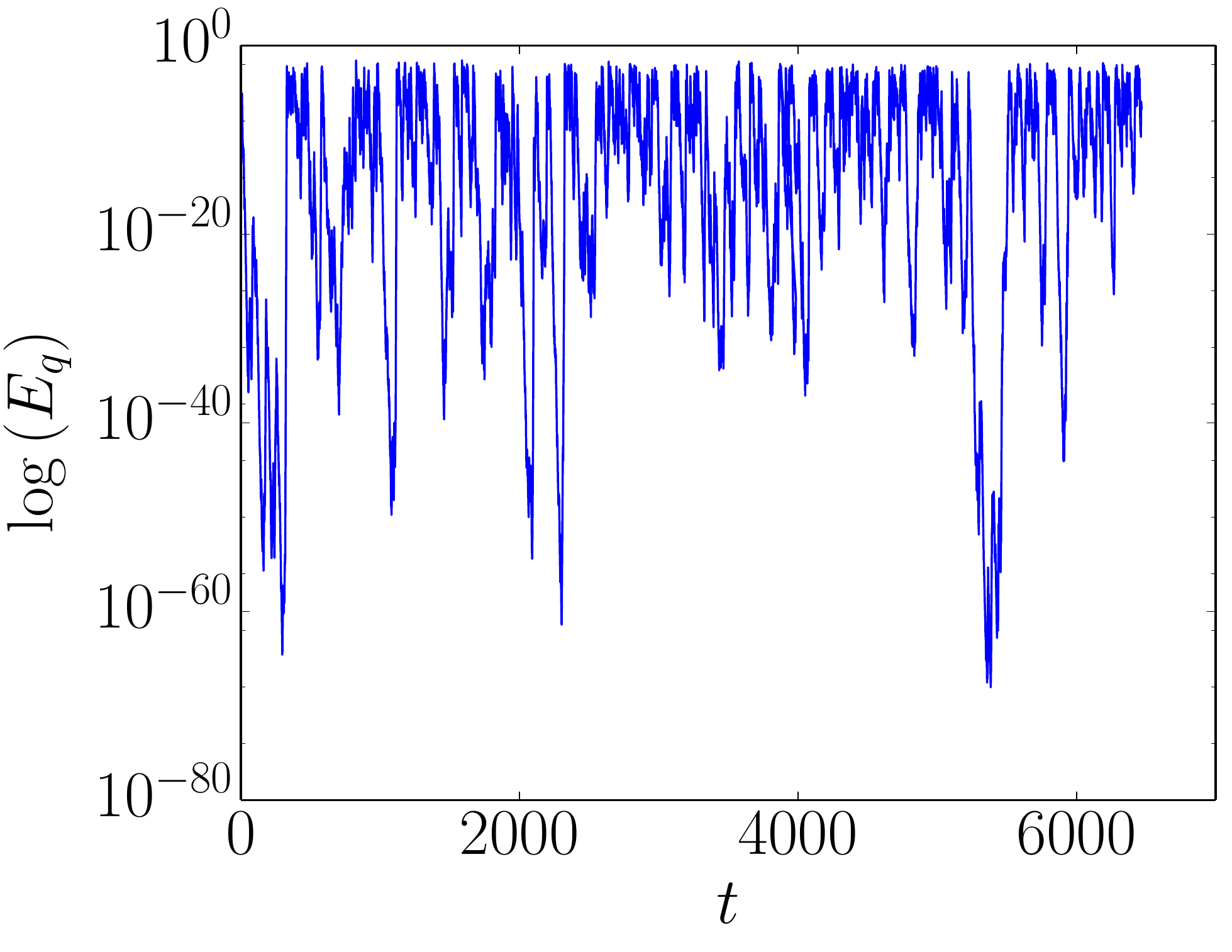}             
    \end{subfigure}                                                        
    \caption{A typical time series of the 3D energy for a $Q$ very close   
    to $\QDC$. The one on the left has linear scalings                     
    and shows the typical picture of an on-off intermittent signal,        
    whereas the right plot shows the same signal but in log-linear scale.} 
    \label{ON-OFF}                                                         
\end{figure}                                                               
Figure \ref{ON-OFF} shows a typical signal for the evolution of the energy of the 3D field $\uq$ as a function  of time,
in linear (left panel) and log-linear (right panel) scale. The signal exhibits bursts of energy followed by times 
with very weak energy. This is a signature of  `on-off' intermittency. 
On-off intermittency is a generic behavior that
appears in the vicinity of an instability in the presence of multiplicative noise \citep{onoff,onoff2}. 
It has been observed in turbulent dynamo simulations \citep{alexakis2008effect}, in electronic circuits \citep{hammer1994experimental}, in electrohydrodynamic convection \citep{john1999off}
and spin-wave systems \citep{rodelsperger1995off}.
%
%
In such situations the unstable modes have a growth rate that varies strongly in time, taking both positive and negative values. 
In its simplest form  on-off intermittency of an unstable mode $X$ is modeled by 
 \beq
 \dot{X} = (\alpha + \xi) X -X^3 \label{xi} 
 \eeq
where $\alpha$ represents mean growth rate and $\xi$ represents a zero mean random noise.
In the thin layer turbulent system
%
the 3D instabilities of the 2D flow,
whose energy is described by Eq. \ref{hami},
%
have an averaged growth rate $\alpha$ that is proportional to 
the deviation from criticality $\alpha \propto(\QDC-Q)$.
At the same time the role of the random  
multiplicative
noise is played by the turbulent fluctuations of the 2D field $\uo$.
If the averaged growth rate $\alpha$ in Eq. \ref{xi} is sufficiently smaller than the fluctuations $\xi$ of the instantaneous growth rate, then 
the system spends long intervals of time with very small amplitudes (off-phase) intervened by short burst where 
the nonlinearities are effective (on-phases). In log scale the amplitude of the unstable mode follows a biased random walk
bounded from above by the nonlinearities. 
The model predicts that, close to the onset, the probability distribution function (PDF) $P(E_{q})$ of the energy of the mode $E_{q} \sim X^2$ follows the scaling
\begin{equation} \label{PDF_on-off}
P(E_{q}) \sim E_{q}^{\alpha /D - 1}, \mathrm{for } \quad X\ll1
\end{equation}
where $D$ is the amplitude of the noise.
Furthermore, the duration of the off-phases $T_{off}$ diverges with the deviation from the onset (here $\QDC-Q$) as as $T_{off} \propto (\QDC-Q)^{-1} $ 
while the amplitude 
$E_{on}$ 
and the time $T_{on}$ in the on-phase becomes independent of ($\QDC-Q$). This implies that the time averaged energy scales like
\beq
\la \mathcal{E}_{q} \ra_{_T} \propto \frac{ E_{on} T_{on} }{T_{off} + T_{on}}  \propto \QDC-Q,
\eeq
where $\la \mathcal{E}_{q} \ra_{_T}$ is also the time average of $E_{q}$. 
Note that this is not the behavior we observe in Fig. \ref{Q2D_Fig}, which suggests a scaling closer to $\la \mathcal{E}_{q} \ra_{_T} \propto (\QDC-Q)^{2}$, 
since the time-averaged $\eqp$ is proportional to the time-averaged of $E_{q}$.
%
\begin{figure}                                                                                                   
  \begin{center}\vspace{0cm}                                                                                     
    \includegraphics[width=0.45\textwidth]{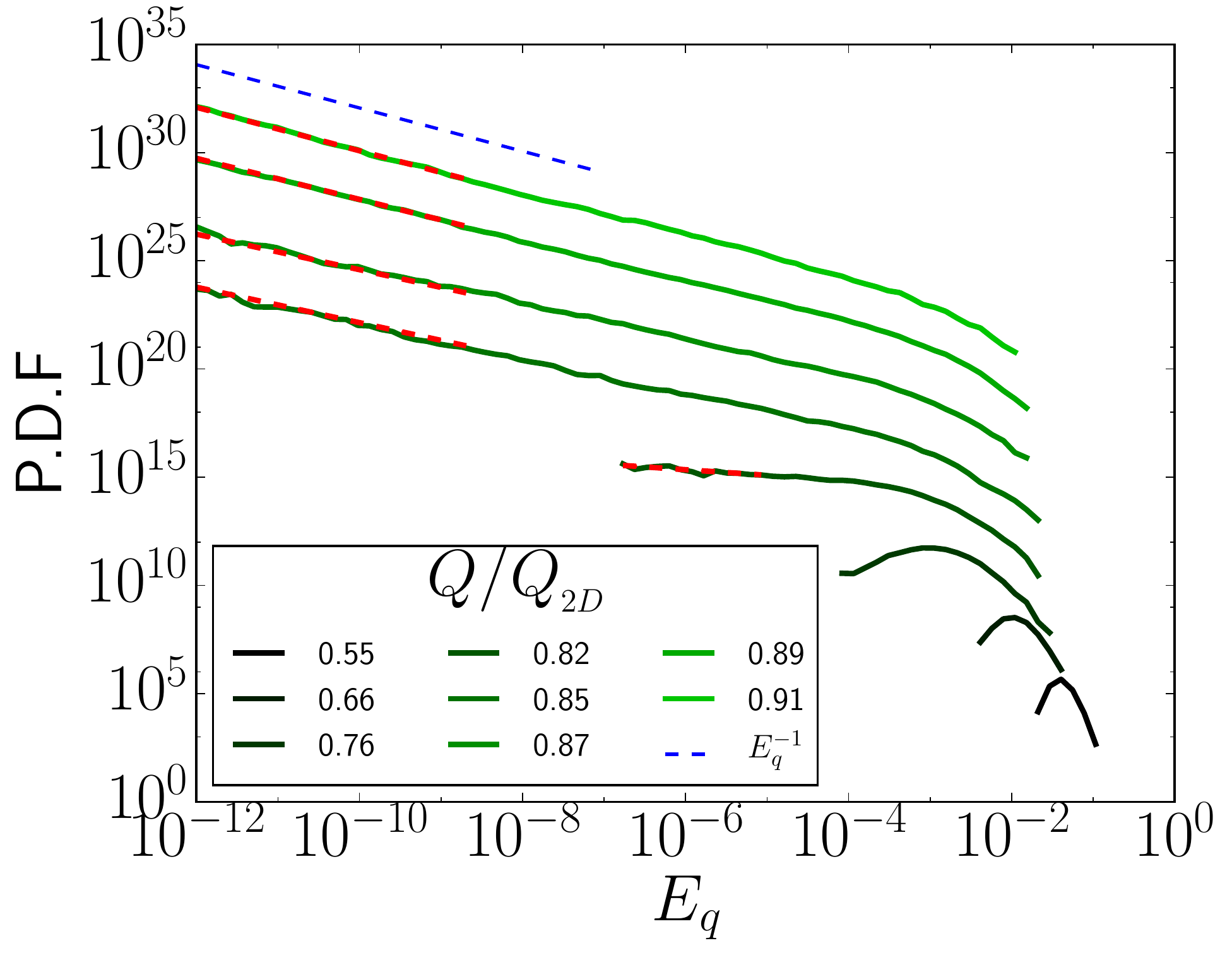}\vspace{0cm}                                      
    \includegraphics[width=0.48\textwidth]{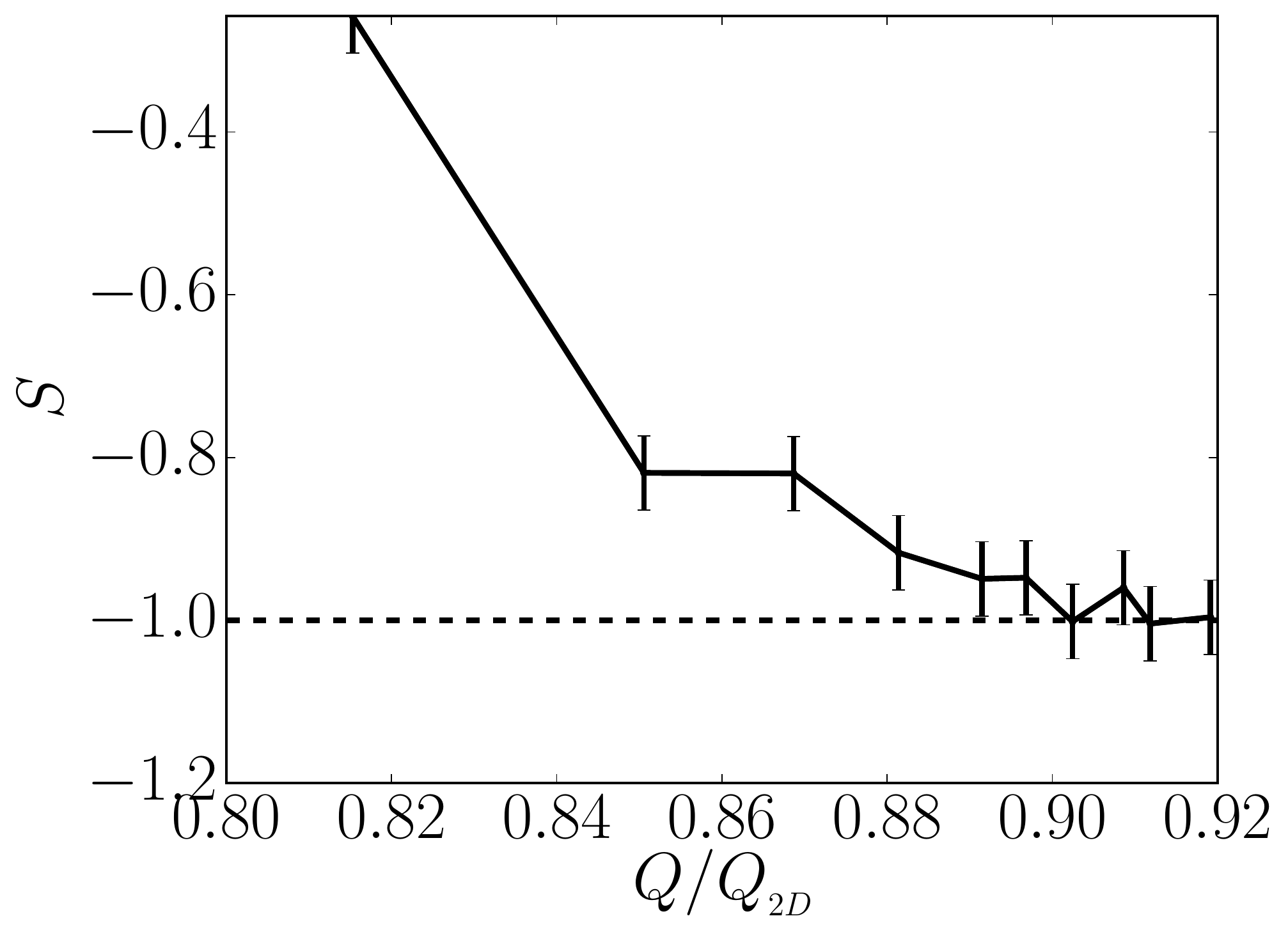}\vspace{0cm}                                      
	  \caption{{\bf Left: } The PDF in time of $E_{q}$. As before, dark means farther away from $\QDC$,            
	  and thus smaller $Q$, and light green is the most 2D (closest to $\QDC$). The blue dashed line               
	  represents $E_{q}^{- 1}$ and the red dashed lines represent where the power law exponent measurements        
	  were taken (if any).                                                                                         
	  {\bf Right:} Various values of the exponents $S(Q)$ versus the fraction                                      
	  $Q / \QDC$ that measures the deviation from criticality.                                                     
	  These values are the slopes of the red dashed lines                                                          
	  in the left panel. Notice the approach to $-1$, as the model predicts.}                                      
	\label{PDF_T}                                                                                                  
  \end{center}                                                                                                   
\end{figure}                                                                                                     
Figure \ref{PDF_T} shows the PDF of the energy $E_{q}$ calculated for different values of $Q$. 
In agreement with the model when the critical value $\QDC$ is approached the PDF becomes singular
showing a power-law behavior $P(E_{q})\propto (E_{q})^{S(Q)}$ . The exponents $S(Q)$  of this power law are 
shown in the right panel of the same figure. As criticality is approached $Q\to\QDC$ these exponents 
tend to $-1$ in agreement again with the model. However this asymptotic value is not approached linearly (i.e. $S(Q) \simeq  (\QDC-Q)/D -1$)
as the model suggests but closer to a quadratic behavior $S(Q) \simeq  (\QDC-Q)^2/D^2 -1 $.  
Thus, just like with the scaling observed in Eq. \ref{sqr} there is a disagreement with this model.  

Resolution comes from looking at the spatial behavior of the unstable modes. 
\begin{figure} 
    \centering                                                                      
    \begin{subfigure}[b]{0.45\textwidth}                                            
        \includegraphics[width=7.0cm]{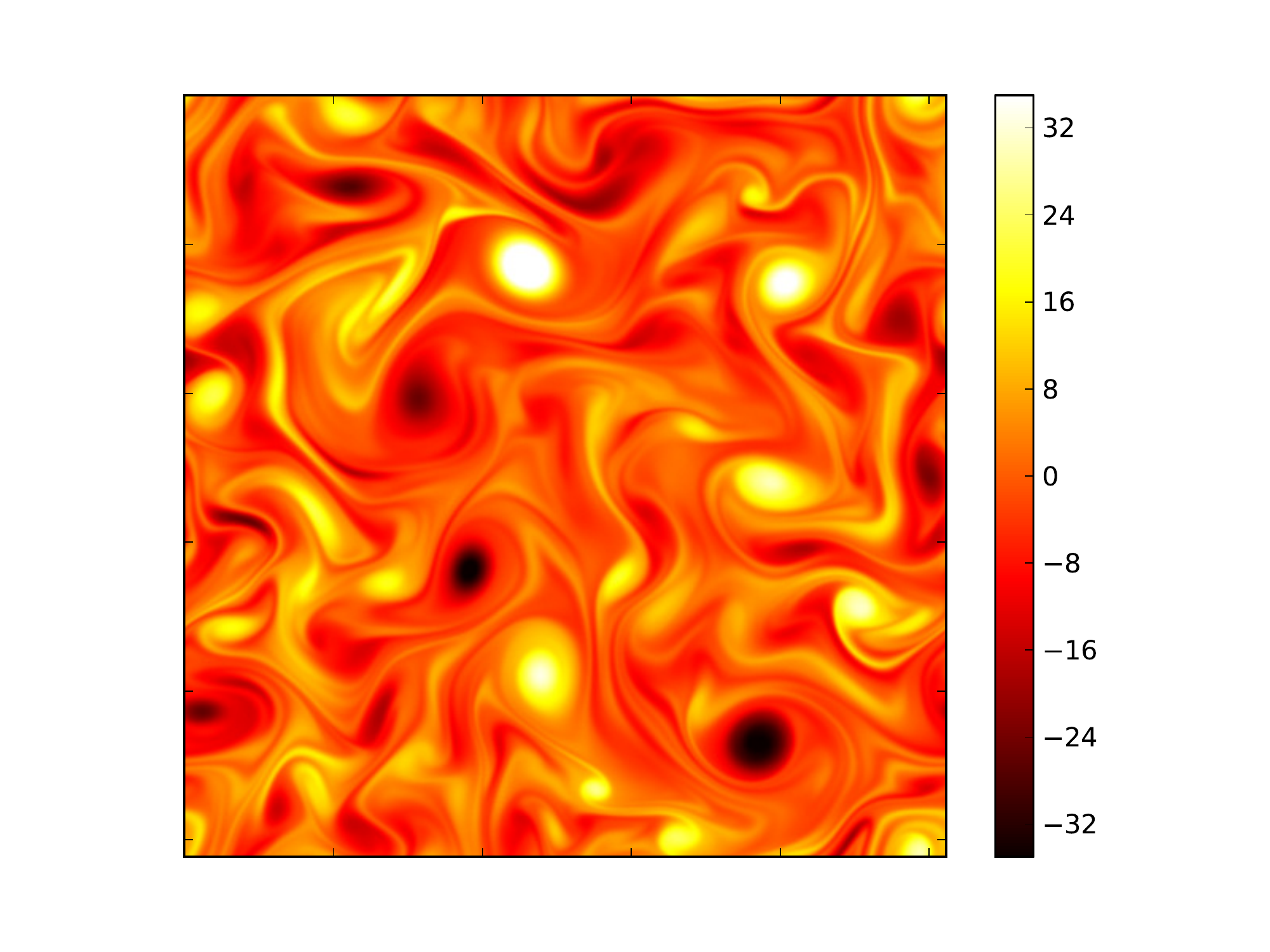}                          
    \end{subfigure}                                                                 
    \begin{subfigure}[b]{0.45\textwidth}                                            
        \includegraphics[width=7.0cm]{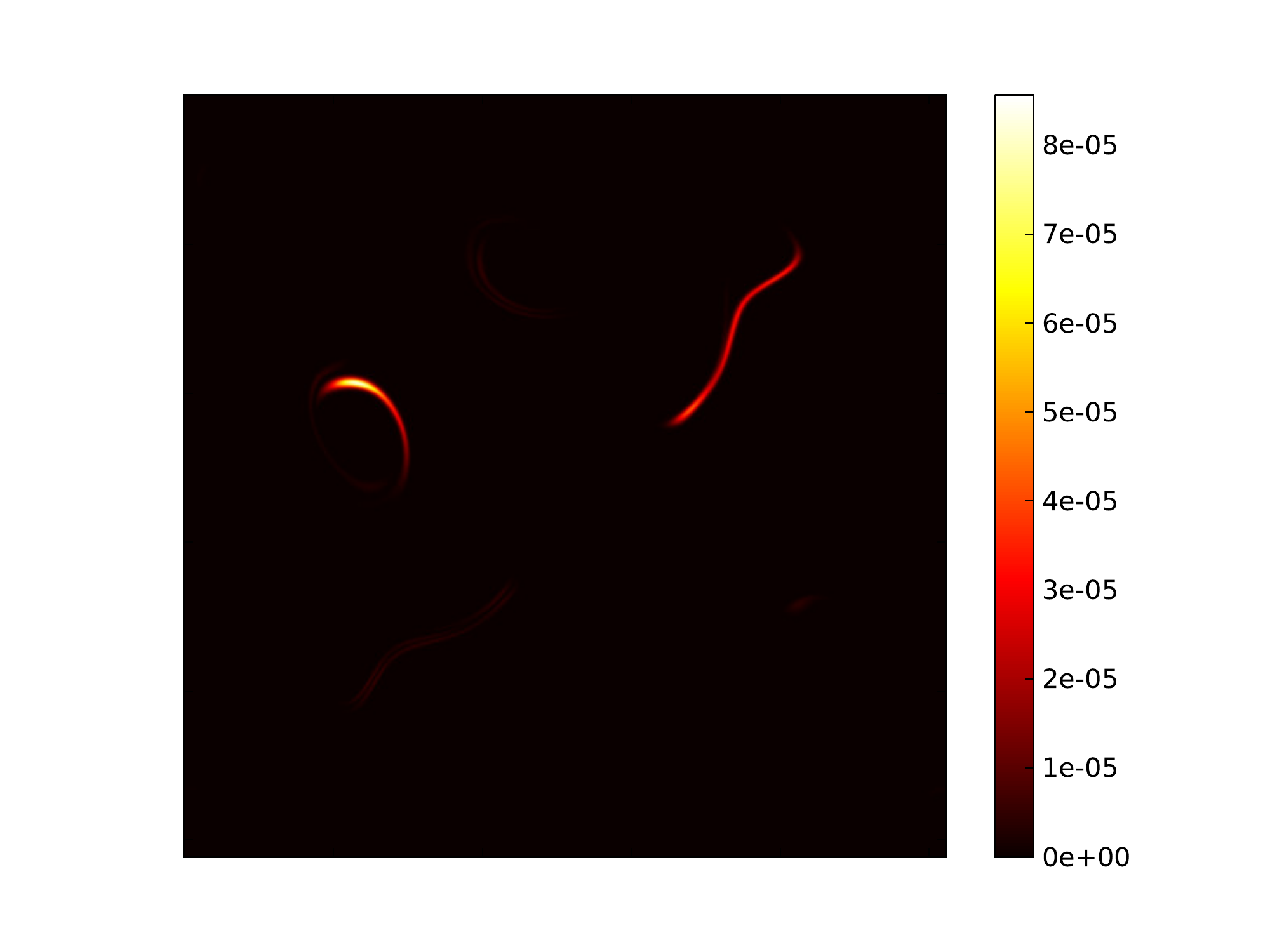}                          
    \end{subfigure}                                                                 
    \caption{The vorticity $\nabla \times \uo$ (left) and the 3D energy density     
    $\mathcal{E}_{q}$ (right)                                                       
    for a point $Q / \QDC = 0.92$, right before $\QDC$.       
    For $k_f L = 8$, and                                                            
    constant forcing amplitude. Note  the very localized structure                  
    of the 3D energy, an extreme of a trend seen in Fig. \ref{prettypix_3DEn}. }    
    \label{prettypix_Q2D}                                                           
\end{figure}                                                                        
Figure \ref{prettypix_Q2D} shows a snapshot of the vorticity $\nabla \times \uo$ on the left panel and of the 3D energy density $\mathcal{E}_q = |\vq|^2 / 4$ in the right panel. 
The vorticity shows the classical  behavior of 2D turbulence. The 3D energy, however, shows a very intermittent behavior in space: most of the
3D energy is concentrated in a single structure occupying a small fraction of the area of the domain size.
Comparing with the bottom panels in Fig. \ref{prettypix_3DEn} we see that, as $Q$ is increased and $\QDC$ is approached, there are
less and less 3D structures occupying a smaller and smaller fraction of the domain area. For $Q$ really close to $\QDC$ we see at the right panel of 
\ref{prettypix_Q2D} that only a single structure or two exist.
The unstable solution thus is not only intermittent in time, but it is also intermittent in space! 
\begin{figure}                                                                                                  
  \begin{center}\vspace{0cm}                                                                                    
    \includegraphics[width=0.45\textwidth]{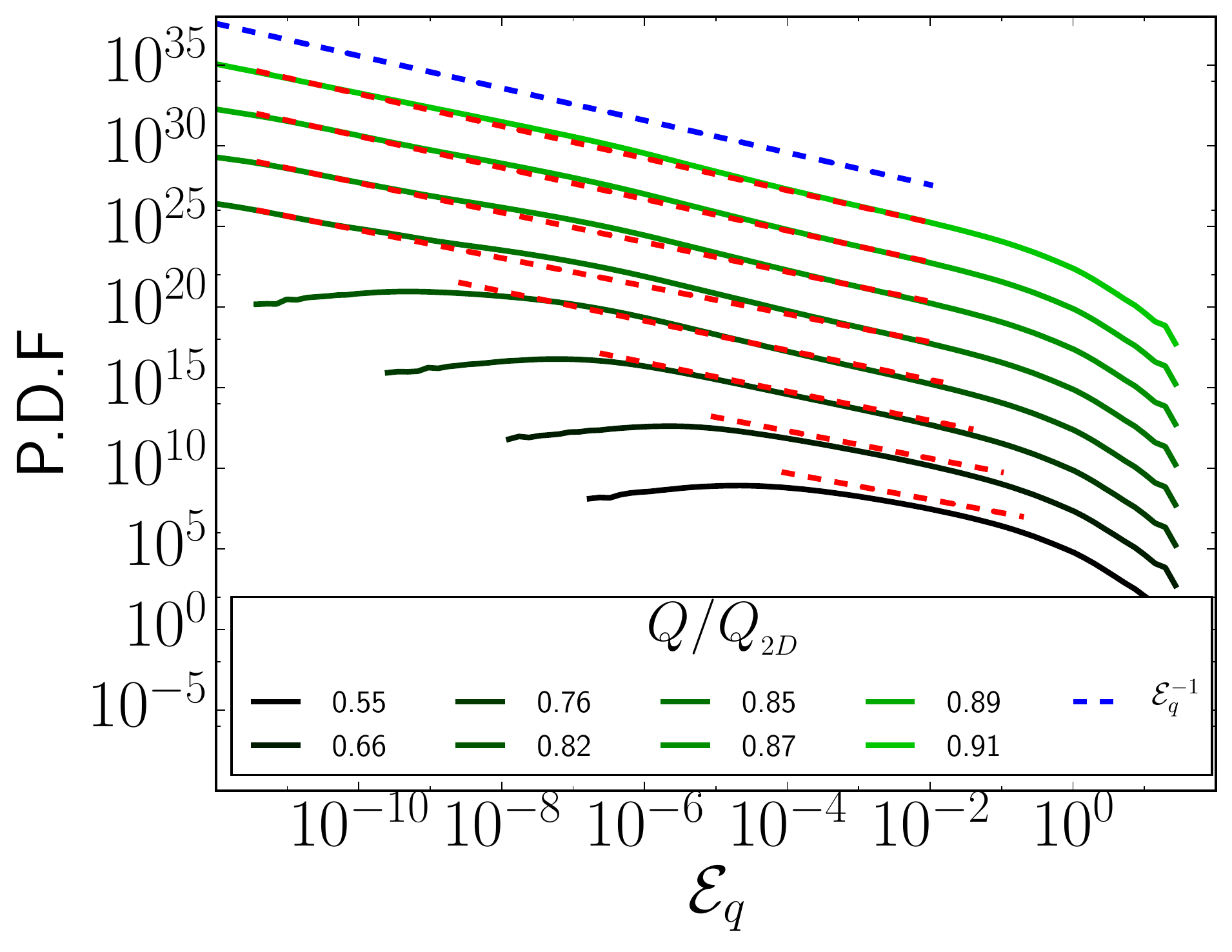}\vspace{0cm}                                     
    \includegraphics[width=0.50\textwidth]{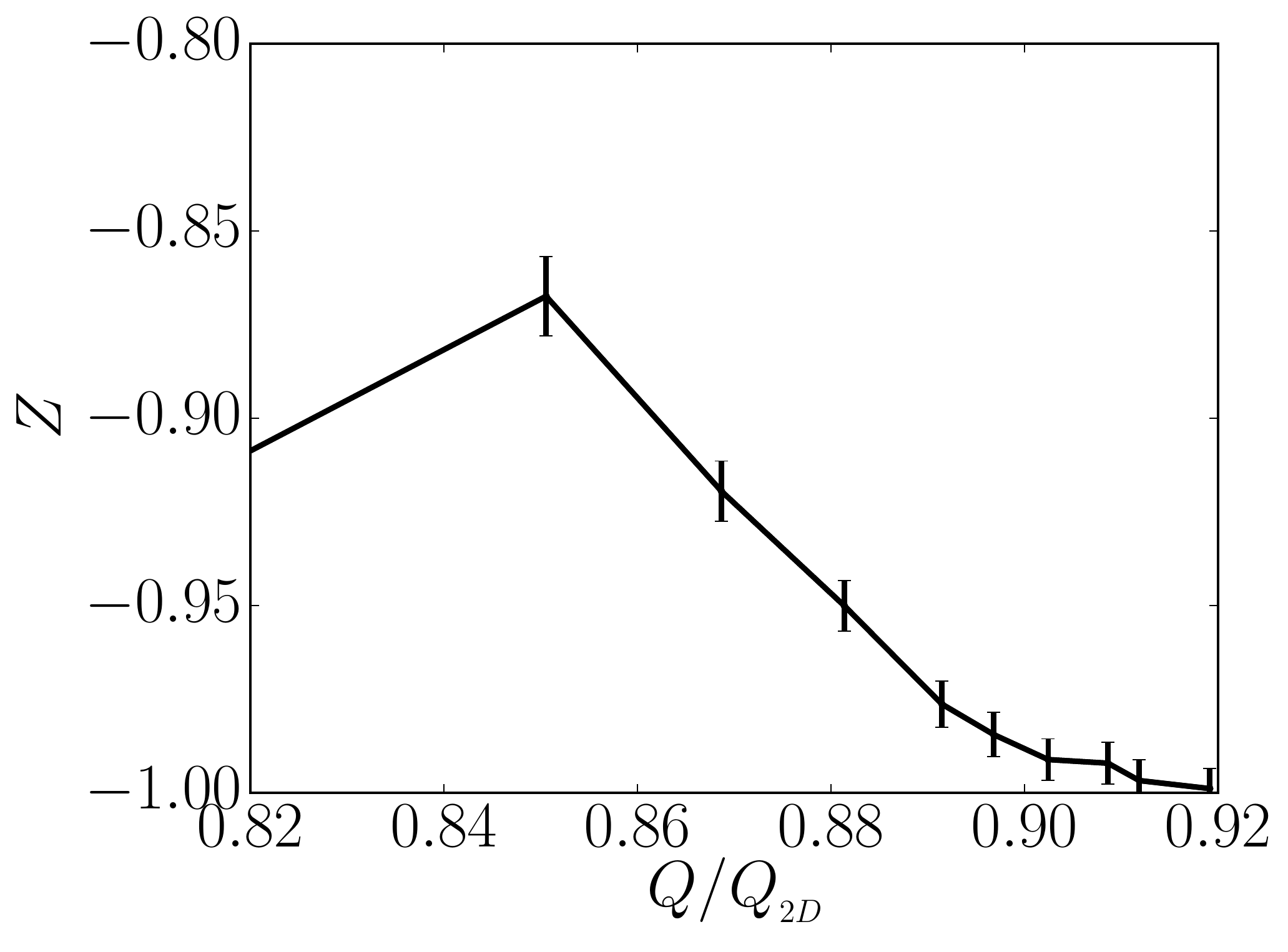}\vspace{0cm}                                     
	  \caption{ {\bf Left: }The PDF in space of $\mathcal{E}_{q}$. As before, dark means farther away from $\QDC$,
	  and thus smaller $Q$, and light green is the most 2D (closest to $\QDC$). The blue dashed line represents   
	  $E_{q}^{- 1}$ and the red dashed lines represent where the measurements of the exponents                    
	  were taken.                                                                                                 
	  {\bf Right:} Various values of the measured power-law exponents $Z$ versus the fraction $Q / \QDC$.         
	  The measured power-laws are demonstrated by the red dashed lines                                            
	  in the left panel.}                                                                                         
	\label{PDF_S}                                                                                                 
  \end{center}                                                                                                  
\end{figure}                                                                                                    

Figure \ref{PDF_S} shows the PDF in space of $\mathcal{E}_{q}$ measured for different values of $Q$. Just like the PDF of the energy $E_{q}$ 
the PDF of  $ \mathcal{E}_{q} $ shows a power-law behavior: as criticality is approached  $P(\mathcal{E}_{q})\propto \mathcal{E}_{q}^Z$ with $Z\to-1$
%
as $Q\to \QDC$.  
Spatial intermittency is not taken in to account in the on-off intermittency model \ref{xi} that takes in to account the evolution of only one single mode. 
In our system the non-linearity (on-phase) is not only visited rarely in time but also rarely in space. 
The averaged energy in space and time will then satisfy 
\beq
\la \mathcal{E}_{q}^n \ra_{_T} = \frac{ \mathcal{E}_{on}^n T_{on} V_{on } }{ (T_{on}+T_{off}) V },
\eeq
where $V_{on}$ is the area occupied by the ``3D-active'' regions, and $V$ is the total area of our system.
%
The scaling of $\la \mathcal{E}_{q}^n \ra_{_T}$ on $(\QDC-Q)$ will thus not only depend on the scaling of the time fraction  $T_{on}/T_{off}$ 
but also on the area fraction $V_{on}/V_{off}$.   
The model in eq. \ref{xi} is not sufficient to describe our system and most likely an extended system with random multiplicative noise both in space
and time will be required to capture correctly the statistics of our system \citep{grinstein1996phase,Horsthemke2006}. Such a possibility will be examined in future work. 

\section{Conclusions}\label{sec:Conc}

In this work we investigated turbulence in a thin layer using numerical simulations of a two-dimensional model of the Navier-Stokes equation
obtained by a severe Galerkin  truncation in the vertical direction. 
The decreased dimensionality of our system allowed us to systematically investigate the transition from a forward to an inverse cascade.
Our results demonstrate the existence of two critical heights (quantified by the parameter $Q$)
with an unexpectedly rich behavior close to criticality. 

The first critical height $H_{_{3D}}=\ell_f/\QTC$ marks the transition from a forward cascade 
for $H > H_{_{3D}}$ to a bidirectional cascade for $H < H_{_{3D}}$. 
Above this critical height $H > H_{_{3D}}$ the 3D component of the flow is in equipartition with the 2D part of the flow
and most of the energy is in the small scales.
Although there is some transfer of energy from $\uo$ to the large scales it is compensated
by the forward transfer caused by the three dimensional $\uq$ field leading to zero net transfer of energy to the large scales.
The spectrum displays a Kolmogorov spectrum at the large wavenumbers.
The structures in this case are small scale vortex tubes occupying a finite fraction of the computational domain.

Below but close to the critical height $H \lesssim H_{_{3D}}$ a weak inverse cascade is observed.
The amplitude of the inverse cascade displays a close to linear dependence with the deviation from criticality $Q-\QTC$ (or $ H_{_{3D}}-H$). 
The transfer of energy to the large scales from the two dimensional field $\uo$ can not be compensated by the forward transfer of
the three dimensional field $\uq$, which is weaker, leading to the observed inverse cascade. The spectrum is close to Kolmogorov $k^{-5/3}$ both  in
the large and the small scales. In real space one observes the coexistence of large scale 2D vortices (similar to a pure 2D flow)
along with 3D vortex tubes. The two distinct structures occupy different regions of space. The fraction of the area they occupy
depends on the deviation from the onset. It is possible that the interactions of these 2D-vortices with the 3D-structures display
predator-prey dynamics, as it has been recently claimed for the transition to turbulence in Couette and Poiseuille flow 
\citep{barkley2015rise,goldenfeld2016turbulence,lemoult2016directed,sano2016universal}, 
and thus this transition could also fall in the universality class of directed percolation \citep{obukhov1980problem}
as suggested by \cite{Pomeau} for sub-critical instabilities in turbulence. 

The second critical height $H_{_{2D}}=\ell_f/\QDC$ marks the transition 
from the bidirectional cascade for $ H_{_{2D}} < H < H_{_{3D}}$ to an inverse cascade for $H < H_{_{2D}}$. 
The critical point $H_{_{2D}}$ is shown to scale like $H_{_{2D}}\propto \ell_f Re^{-1/2}$. It can be shown that
for all $H < H_{_{2D}}$ all 3D perturbations decay exponentially in time. The value of $H_{_{2D}}$ was estimated
by rigorous bounds. For values  of $H$ larger but close to  $H_{_{2D}}$ the three dimensional flow exhibited 
a strongly intermittent behavior. The total energy displayed on-off intermittency behavior in time with bursts of energy.
At the same time intermittent behavior was also observed in space with the 3D vortex tube like structures occupying 
lesser domain area the closer $H$ is to $H_{_{2D}}$. This intermittent behavior results from 
the almost linear evolution of the unstable 3D mode $\uq$  driven by the spatio-temporal fluctuations of the 
2D turbulence. The transition close to this point then could possibly modeled by extended systems in the presence of multiplicative noise
\citep{grinstein1996phase,Horsthemke2006}.

The precise statistical description and the possible universality class of the two critical points requires certainly further investigation.
The present investigation however clearly demonstrated the non-triviality of the two critical points and their  unexpectedly rich behavior.
Whether similar transitions are observed in other systems like rotating, stratified or magnetized flows remains to be seen.

\bibliographystyle{jfm}
\bibliography{thin_layer}

\end{document}